\documentclass[aps,prd,nofootinbib]{revtex4}
\usepackage{amssymb,epsfig,color,graphicx}                  
\newcommand{\LCDM}{$\Lambda$CDM}

\begin{document}
\title{Findings of the\\ Joint Dark Energy Mission \\ Figure of Merit
Science Working Group}

\author{Andreas~Albrecht, Luca~Amendola, Gary~Bernstein, Douglas~Clowe,
Daniel~Eisenstein, Luigi~Guzzo, Christopher~Hirata, Dragan~Huterer,
Robert~Kirshner, Edward~Kolb, Robert~Nichol}

\begin{abstract}

These are the findings of the Joint Dark Energy Mission (JDEM) Figure of Merit
(FoM) Science Working Group (SWG), the FoMSWG.  JDEM is a space mission planned
by NASA and the DOE for launch in the 2016 time frame.  The primary mission is
to explore the nature of dark energy.  In planning such a mission, it is
necessary to have some idea of knowledge of dark energy in 2016, and a way to
quantify the performance of the mission.  In this paper we discuss these
issues.  

\end{abstract}

\date{Dec 7, 2008}

\maketitle

\section{The Unknown Nature of Dark Energy}

The discovery that the universe is expanding with an ever-increasing velocity is
now a decade old, yet there is no compelling theoretical explanation. We have a
cosmological standard model, called $\Lambda$CDM, that seems capable of
accounting for (at least in principle) all cosmological observations, including
the apparent acceleration.  But it is sobering to note that in $\Lambda$CDM as
much as 95\% of the present mass-energy of the universe is not understood, with
only 5\% of the present mass-energy in the form of ``stuff'' we understand
(baryons, radiation, neutrinos).  The rest of the present mass-energy of the
universe is assumed to be dark: about 30\% in the form of dark matter providing
the bulk of the gravitational binding energy of galaxies, galaxy clusters, and
other large-scale structure, and about 70\% in the form of dark energy driving
the present expansion of the universe. Both dark matter and dark energy point to
physics beyond the standard models of gravity or particle physics.  

This paper is concerned with dark energy \cite{REVIEWS}, the {\it primum
mobile} for the present accelerated expansion of the universe.

While $\Lambda$CDM seems capable of accounting for all observations, the aim of
cosmology is not simply to find a model that describes the observations, but
rather to find one that agrees with observations {\it and} is also grounded in
physical reality.\footnote{Cosmological models that describe observations but
are not grounded in physical reality have been found in the past, but have been
rejected in favor of models based on the laws of nature (see, e.g.,
\cite{Ptolemy}).}  The most important task ahead is to discover the nature of
the dark universe, in particular, dark energy.

To date, all indications of dark energy come from measuring the time evolution
of the expansion history of the universe.  In the standard
Friedmann-Lema\^{i}tre-Robertson-Walker (FLRW) cosmology, the expansion rate as
a function of the scale factor $a$ is given by the Friedmann 
equation\footnote{The scale factor $a$ is normalized to unity at present.  
It is related to the redshift $z$ by $1+z=1/a$.} 
\begin{equation}
H^2(a) = H_0^2\left[\Omega_R a^{-4} + \Omega_M a^{-3} 
       + \Omega_k a^{-2} + \Omega_{DE}
       \exp\left\{3\int_a^1 \frac{da'}{a'}\left[1+w(a')\right] \right\} \right].
\end{equation}
In this expression $\Omega_i$ is the present fraction of the critical density,
$\rho_C=3H^2_0/8\pi G$, in the form of component $i$; e.g., radiation ($R$),
matter ($M$), curvature ($k$) and dark energy ($DE$).  The parameter $H_0$ is
the present value of the expansion rate of the universe (Hubble's constant).
Finally, $w(a)$ is the ratio of the pressure to the energy density for dark
energy, $w(a)=p(a)/\rho(a)$.  If dark energy is Einstein's cosmological
constant, $w(a) = -1$.

In framing the question of the nature of dark energy, it is useful to start
with something that doesn't work: It is clear from the observations that the
Einstein--de Sitter cosmological model (a spatially flat, matter-dominated,
FLRW model) does not describe the recent expansion history of the universe.  In
FLRW models the Friedmann equation follows directly  from the $0-0$ component
of the Einstein equations, so the fact that the Einstein--de Sitter model fails
can be expressed as 
\begin{equation} 
\label{neq}
G_{00}(\textrm{spatially flat FLRW}) \neq 8\pi G T_{00}(\textrm{matter}) .
\end{equation}

There are two generally orthogonal directions in explaining the observations. 
The first direction is to assume there is, in addition to matter and radiation,
a new type of ``negative pressure'' component to the energy density of the
universe that would be added to the right-hand-side of Eq.\ (\ref{neq}).  The
other direction is modify the left-hand side of Einstein's equation by saying
that either the metric used is inappropriate, or to interpret the failure of
Eq.\ (\ref{neq}) as an indication that Einstein did not have the final word on
gravity, and that there is effectively a modification of the left-hand-side of
his equations.

\subsection{Primary Dark Energy (the Right-Hand-Side of Einstein's equations)}

Even before the observation of cosmic acceleration, physicists were familiar
with two possible sources of cosmic  acceleration. These two types of
acceleration were already well established, to the point that they were
discussed in basic textbooks. And  both can be thought of as a new fluid that
contributes to the stress tensor.  In one case the density of the new fluid is
constant (equivalent to Einstein's ``cosmological constant''), and in the other
the fluid density is dynamical (``quintessence''). 

If cosmic acceleration is driven by a new fluid, the first question is to
discriminate between these two ideas by determining whether the fluid density is
constant or dynamical.

\subsubsection{Einstein's Cosmological Constant}

The cosmological constant arises as an additional constant term in  Einstein's
equations. A FLRW universe with a cosmological constant with value $\Lambda$ is
equivalent to adding a matter component with density $\rho_\Lambda \equiv
\Lambda / 8\pi G$ and with equation of state $p =-\rho$ (or $w(a) = -1$). This
equation of state describes a $\rho_\Lambda$ that is constant throughout the
evolution of the universe.  Until the mid 1990's, a commonly held belief among
particle physicists and cosmologists was that $\Lambda$ was identically zero. 
For many, this belief was partly motivated by the fact that na\"{\i}ve
calculations give quantum field contributions to $\Lambda$ of order $10^{60}$ or
$10^{120}$ times larger than observationally acceptable.  Other possible
contributions to the vacuum energy, like scalar field condensates, are also
truly enormous compared to the allowed value.  It was widely believed that
finding some symmetry or dynamical process that sets $\Lambda$ (including
contributions from quantum fields) precisely to zero was the best hope to
resolve this apparent discrepancy.  

The observation of cosmic acceleration has forced radical changes to this
thinking. Many theorists now favor a ``string theory landscape''
\cite{Bousso:2000xa,Kachru:2003aw,Bousso:2007gp}  where all possible values of
$\Lambda$ are represented in a ``multiverse.''  Cosmology and particle physics
truly become one in this picture, since what we see as our ``fundamental''
particles and forces would be dictated by our cosmological evolution around the
landscape. Alternatively a single ``pure'' cosmological constant might be what
nature has chosen, but that would lead to dramatic changes to how we think about
cosmology, entropy, and ``heat death'' in our universe
\cite{Dyson:2002pf,Albrecht:2004ke}.

A precision experiment supporting $w=-1$ would lend weight to these radical
ideas, while any evidence for $w\neq-1$ would falsify most of them instantly. 

\subsubsection{Quintessence Models}

Another widely held belief that has had growing support over the last few
decades is that the universe underwent a period of {\em cosmic inflation} in the
distant past.  A period of cosmic inflation  appears to explain many features
observed in the universe today, some of which  seemed puzzling before the idea
of inflation.  

Today, inflation is quite well understood in terms of its phenomenology, but it
still has a number of unresolved foundational questions. Despite these, it is
clear that cosmic inflation requires a period of cosmic acceleration that {\em
cannot} be described by a cosmological constant.  Cosmic inflation is understood
to be driven by some matter field, typically called the ``inflaton,'' which
exhibits an equation of state $p = w\rho$.  During inflation, $w$ takes values
that approach $w=-1$.  Inflation is fundamentally a dynamical process: Not only
must the universe enter and exit inflation (leading to large variations in $w$)
but small deviations from $w=-1$ are also necessary throughout the inflationary
period in order for the mechanisms of inflation to work properly.\footnote{One
probe of inflation is the search for gravitational waves produced during the
inflationary epoch (for example through their impact on CMB polarization or
directly using gravitational wave detectors).  For common types of inflation
such observations are equivalent to probing the deviations from $w(a)=-1$ during
the inflationary epoch.} Thus, to the extent one believes in cosmic inflation,
one believes the universe has already realized in the distant past cosmic
acceleration using a {\it dynamical} fluid.

Most dynamical theories of dark energy re-use many of the ``rolling scalar
field'' ideas from inflation and go under the name ``quintessence''
\cite{REVIEWS}.  Unfortunately, basic questions about how a quintessence
field with the right sorts of couplings fits in a  fundamental theory are in no
better shape than our understanding of the inflaton.  Still, the prominence of
cosmic inflation in our understanding of modern cosmology, our long history of
preferring $\Lambda = 0$ (see for example \cite{Efstathiou:1990xe}) and  the
presence of several quintessence models with interesting motivations have
generated considerable interest in applying these dynamical ideas to the
acceleration. Furthermore, recent work has shown that results from
advanced experiments could potentially rule out many or perhaps even
most quintessence models entirely \cite{Barnard:2008mn}.   

\subsection{Modification of General  Relativity (the Left-Hand Side of 
Einstein's Equations)} 

The Hubble diagram of Type Ia Supernovae implies the existence of dark energy
only if we assume that the standard cosmological framework in which we interpret
the observations is correct. Part of this basis is Einstein's General Theory of
Relativity (GR), our current standard theory of gravity.  In fact, instead of
adding an extra term (dark energy) to the stress-energy tensor on the right side
of Einstein's equations, one can devise modifications of the left side that
reproduce the observations equally well.  Attempts in this direction include
higher-order curvature terms (as in so-called $f(R)$ theories, \cite{fr}), or
higher-dimensional theories, as in {\it braneworld} models \cite{braneworld}. 
Clearly, this alternative explanation would have consequences as profound and
revolutionary as the existence of dark energy.  Either dark energy or modified
gravity would point to physics beyond our standard models of particle physics
and/or gravity.   

Therefore, we feel that while the accurate characterization of the expansion
history of the  universe, $H(z)$, (which translates into an effective dark
energy equation of state $w(z)$) is crucial for future experiments, it cannot be
considered as the final word.  The ``acceleration problem'' reduces to this
simple quest for one function only if GR is the correct theory of gravity (and
if the dark energy is very simple, for example a scalar field, i.e.,
quintessence).  In general, methods providing independent measurements of
distance and redshift probe the expansion history $H(z)$ (thus $w(z)$), but
cannot tell whether this comes from a true extra fluid in the cosmic budget, or
by a more elusive change in the laws of gravity.  

Luckily, there may be an alternative way to tackle this problem. In addition to
governing the overall expansion of the homogeneous universe, gravity is also
responsible for the gradual build-up of the large-scale structure of the
universe we observe. The efficiency of this mechanism at different epochs, i.e.,
the ``growth rate'' of density fluctuations $f(z)$, depends not only on how fast
the universe expands (thus on $H(z)$), but also on the very nature of the
gravitational force.  Measurements of $f(z)$ (or its integral, the ``growth
factor'' $G(z)$) can break the above degeneracy between simple dark energy
models and modified gravity. 

The growth factor can be obtained from different observables.  When matter
density fluctuations have low amplitude, in the so-called linear regime of
growth, their evolution on sub-horizon scales is easy to describe through a
second-order differential equation, $\ddot{\delta}+2H\dot{\delta}=4\pi {\rm
G}\rho\delta$.  Here  $\delta$ is the relative density excess with respect to
the mean density of matter $\rho$.  The source term on the right side of this
equation depends directly on the theory of gravity: the form used here is
correct only if GR is correct (and if one makes the very reasonable assumption
that primary dark energy, if it exists, is not clustered on these scales). 
This equation has a growing solution such that we can write it as the product
of a spatial part $D(\vec x)$ and a time-dependent part $G(a)$ [where time is
here parametrized via the cosmic scale factor $a(t)$, related to the redshift
by $1+z=1/a(t)$].  Again, $G(a)$ is  the growth factor and its derivative
$f(a)= d\ln(G)/d\ln(a)$ is the growth rate.   Given a measured $H(a)$ (which
appears on the left-hand-side of the $\ddot{\delta}$ equation) and a theory of
gravity (that determines the right-hand-side), the resulting $G(a)$ is 
uniquely predicted.  A discrepancy in this prediction with a direct measurement
of $G(a)$ would be a strong hint that GR breaks down on very large scales.  

Measurements of galaxy clustering at different epochs provide information on
$G(a)$, once the {\it bias} function mapping mass fluctuations into galaxy
fluctuations is known.  The bias issue can be overcome by {\it weak lensing
tomography}, which probes directly the amplitude of mass fluctuations in
redshift slices.  Galaxy peculiar velocities depend directly on the derivative
of $G(a)$, i.e., on $f(a)$, and can be measured from {\it redshift-space
distortions} in large redshift surveys of galaxies. The growth factor $G(a)$
also enters in determining the number density and evolution of {\it rich
clusters of galaxies} expected above a given mass threshold at a given
redshift. 

In fact, the fully relativistic treatment of the growth of small scalar
perturbations shows that they have to be described in terms of two functions,
$\phi$ and $\psi$, which are a generalization of the Newtonian gravitational
potential. In standard Einstein's gravity with an ordinary dark sector the {\it
anisotropic stress}, defined as $\phi-\psi$, vanishes.  Galaxy clustering and
peculiar velocities depend only on $\psi$, since neither strong fields nor high
velocities are involved in the process.  The dual nature of the potential can be
evidenced only under relativistic conditions. This is the case for {\it
gravitational lensing}, a phenomenon that is not present in the classical theory
of Newton and which is  sensitive to both potentials.  As such, it is virtually
the only experimental technique capable of detecting a non-zero anisotropic
stress, i.e., a distinction between the two potentials (another probe is
provided by the Integrated Sachs-Wolfe effect on the CMB, but with much lower
sensitivity).  Such a finding would be a strong indication of either modified
gravity or new physics in the dark sector. 

As this is a Working Group for a space-based JDEM, we will not discuss other
searches for departures from GR.  We only note here that precision GR tests,
e.g., solar system tests, already severely restrict the possible departures
from GR, although constraints obtained on local scales do not necessarily apply 
to cosmological scales.

The implications of this simple reasoning is that the problem of cosmic
acceleration depends on measuring (a) the expansion history; (b) the
growth history, and (c) the two gravitational potentials.

\subsection{Further Alternatives}

A third possibility is to postulate that the formation of structure at
late times leads to large deviations from the isotropic and homogeneous FLRW
metric. Toy models have been constructed (most of them using a
Lema\^itre-Tolman-Bondi (LTB) metric).  These can explain the data without dark
energy, although at the price of introducing large and so-far unobserved
inhomogeneities. These toy models can be tested, e.g., by combining an angular
diameter distance (like tangential BAOs) with a direct measurement of the
expansion rate $H(z)$ (e.g., from radial BAOs).

\subsection{The Task Before Us}
\label{sec:task}

The standard cosmological model assumes that dark energy is described by a
cosmological constant.  While this is the simplest possibility, the magnitude of
$\Lambda$ is difficult to understand, and, of course, we seek tests of this
hypothesis.  

Thus, when discussing ways to probe dark energy, we will adopt the approach
of the Dark Energy Task Force (DETF) \cite{DETF} and assume that the 
observational program should be to 
\begin{enumerate}
\item determine as well as possible whether the accelerating expansion is
consistent with a cosmological constant, i.e., unevolving dark-energy density,
\item measure as well as possible any time evolution of the dark energy density,
and
\item search for a possible failure of general relativity through comparison of
the effect of dark energy on cosmic expansion with the effect of dark energy on
the growth of cosmological structures.
\end{enumerate}

In this report we establish a framework for assessing progress toward those
goals.  Section \ref{sec:fisher} reviews the Fisher-matrix methodology by which
the power of experiments is described; Section \ref{sec:fiducial} describes the
FoMSWG cosmological model and its free parameters. Section \ref{sec:circa}
describes the knowledge of these parameters that we expect at the advent of the
JDEM era, with details of these pre-JDEM ``data models'' given in the
Appendices.  The FoMSWG finds that 
potential payload implementations and observational strategies
should be judged on their
advancement over this set of standardized pre-JDEM data models. 
Section \ref{sec:foms} describes quantities and plots which we suggest be used
to gauge the effectiveness of proposed JDEM investigations, 
and which can be derived
from the Fisher matrices of JDEM data models.  Section \ref{sec:interp} presents
guidelines for the interpretation of our suggested plots and figures of merit.

We do not discuss the important issue of systematic errors in dark energy
measurements, nor do we discuss the optimization of a dark energy program to
accomplish the three goals above.  
Again, it is important to stress
that our goal is simply to establish a framework for assessing progress.

\section{The Fisher Information Matrix}
\label{sec:fisher}

The FoMSWG adopted a strategy of describing the potential of any given dark
energy
investigation by its Fisher information matrix over the parameters of a
standardized cosmological model.  In this section, we review the meaning of the
Fisher matrix and typical manipulations required to implement the FoMSWG
evaluations of experimental scenarios.  In the following section we describe the
cosmological model adopted by the FoMSWG.

The Fisher matrix method was used by the Dark Energy Task Force and is 
standard in many fields.  It is instructive to consider first a simple case:
suppose we observe a series of quantities $y_b$, $b\in\{1,\ldots B\}$, each of
which has Gaussian uncertainties $\sigma_b$.  Suppose in addition that each
observable should be described by a function $f_b$ of some parameters $p$.  The
common $\chi^2$ value is (we will assume $y_b$ for different $b$ are 
uncorrelated)
\begin{equation}
\chi^2=\sum_{b=1}^B\frac{\left(f_b(p)-y_b\right)^2}{\sigma_b^2}.
\end{equation}
If the parameters $p$ describe the true universe, then the likelihood of a 
given set of observations is
\begin{equation}
P(y) \propto \exp\left(-\frac{1}{2}\chi^2\right).
\end{equation}

The problem, however, is to estimate parameters $p$ given a realization of the
data $y$.  Using Bayes' theorem with uniform prior, we have $P(p|y)\propto
P(y|p)$, so that the likelihood of a parameter estimate can be described as a
Gaussian with the same $\chi^2$, now viewed as a function of parameters.  If we
expand about the true values of the parameters, $p^i=p^i_0+\delta p^i$, and
average over realizations of the data, 
\begin{equation}
\left\langle\chi^2(p)\right\rangle = \left\langle\chi^2\right\rangle + 
\left\langle\frac{\partial\chi^2}{\partial p^j}\right\rangle\delta p^j +
\frac{1}{2}\left\langle\frac{\partial^2\chi^2}{\partial p^j\partial p^k}
\right\rangle\delta p^j\delta p^k + \ldots
\end{equation}
where the expectation values are taken at the true values $p_0$.  The mean
value of observable $y_b$ is indeed $f_b(p_0)$, so the second term
vanishes.  The distribution of errors in the measured parameters is thus in the
limit of high statistics proportional to
\begin{equation}
\exp\left(-\frac{1}{2}\chi^2\right) \propto 
\exp\left(-\frac{1}{4}\left\langle
\frac{\partial^2\chi^2}{\partial p^j\partial p^k}\right\rangle
\delta p^j\delta p^k\right) =
\exp\left(-\frac{1}{2}\mathcal{F}_{jk}\delta p^j\delta p^k\right)
\end{equation}
where the Fisher matrix is
\begin{equation}
\mathcal{F}_{jk} = \sum_b \frac{1}{\sigma_b^2}
\frac{\partial f_b}{\partial p^j} \frac{\partial f_b}{\partial p^k}.
\end{equation}
From this expression it follows that
\begin{equation}
\left\langle\delta p^j \delta p^k\right\rangle
=\left(\mathcal{F}^{-1}\right)^{jk},
\end{equation}
In other words, the covariance matrix is simply the inverse of the Fisher matrix
(and vice versa).

More generally, if one can create a probability $P(p^i|y_b)$ of the model
parameters given a set of observed data, e.g., by Bayesian methods, then one 
can define the Fisher matrix components via \begin{equation} \label{flnp}
\mathcal{F}_{ij} = - \left\langle \frac{\partial^2\ln P}{\partial p^i \partial
p^j}\right\rangle \end{equation} and the Cramer-Rao theorem states that any
unbiased estimator for the parameters will deliver a covariance matrix on the
parameters that is no better than $\mathcal{F}^{-1}$. The Fisher matrix
therefore offers a best-case scenario for one's ability to constrain
cosmological parameters given a set of observations.

The Fisher matrix is useful for planning a suite of experiments, because the
constraints using the ensemble of experiments are described by the {\em sum} of
the Fisher matrices of each component experiment.  This is clear from
$\mathcal{F}\propto \ln P$.  Prior information can also be included the same
way. Furthermore, there are other straightforward manipulations of the Fisher
matrix that yield the expected constraints on subsets or reparametrizations of
the original parameter set.  We offer details on these procedures below.

\subsection{Fisher Matrix Manipulations}

There has been some concern about degenerate or ill-conditioned Fisher
matrices, for instance the Planck prior given by DETF.  In practice,
such (near) degeneracies are an accurate reflection of the inability
to distinguish all the cosmological parameters in a particular
experiment's data.  We need, however, to be careful that round off
error in Fisher-matrix calculations does not significantly change the
constraints implied by the matrix.

One strategy to avoid ill-conditioned matrices is to keep all the
parameter variations at the same order of magnitude.  This is one
motivation for putting all distances into $c/H_{100}$ units and for
using logarithmic quantities for parameters.

A second strategy is to be careful about inverting Fisher matrices, doing so
only when necessary and properly handling the potentially large
covariances that result.  

We now discuss manipulations of Fisher matrices and discuss proper handling
where appropriate.

\subsubsection{Priors}
\label{sec:addprior}

A Gaussian prior with width $\sigma$  can be placed on the $i^{th}$ 
parameter by adding to the appropriate diagonal element of the Fisher matrix:
\begin{equation}
\mathcal{F}_{kl} \longrightarrow \mathcal{F}_{kl} 
+ \frac{\delta_{ki}\delta_{li}}{\sigma^2} \quad\quad(\textrm{no sum on } i),
\end{equation}
which can also be written as
\begin{equation}
\mathcal{F} \longrightarrow \mathcal{F}+\mathcal{F}^p,
\end{equation}
where in this case $\mathcal{F}^p$ is an extremely simple matrix with a 
single  non-zero diagonal element $1/\sigma^2$ in the $i^{th}$ row and
column.

If the prior on $p_i$ is intended to be very strong, the addition of a
large $1/\sigma_i^2$ to $\mathcal{F}$ can cause numerical
instabilities in some algorithms for inversion of the matrix.  An
infinitely strong prior on $p_i$ is equivalent to simply striking the
$i$th row and column from the Fisher matrix.

\subsubsection{Projection onto new variables}
\label{sec:projection}

A common operation is to transform the Fisher matrix in some variable set $p^i$
into a constraint on a new variable set $q^i$.  The formula for this is to
generate the derivative matrix $\mathcal{M}$ with $\mathcal{M}^i_{\ j}=\partial
p^i / \partial q^j$.  In the simple case of Gaussian errors on
observables, the new Fisher matrix is
\begin{equation}
\label{steen}
\mathcal{F}'\,_{lm} = 
\sum_b \frac{1}{\sigma_b^2}\frac{\partial f_b}{\partial q^l}
\frac{\partial f_b}{\partial q^m} =
\sum_b \frac{1}{\sigma_b^2}
\frac{\partial p^j}{\partial q^l}
\frac{\partial p^k}{\partial q^m}
\frac{\partial f_b}{\partial p^j}
\frac{\partial f_b}{\partial p^k} =
\frac{\partial p^j}{\partial q^l}
\frac{\partial p^k}{\partial q^m} \mathcal{F}_{jk} =
\left(\mathcal{M}\right)^T \mathcal{F\,M} 
\end{equation}
using the usual summation convention on $j,k$.  The result holds when
we generalize to arbitrary likelihood functions.

No numerical instabilities are generated by having $\mathcal{F}$ or
$\mathcal{M}$ be rank-deficient, so remapping onto a new variable set will not
generally raise numerical issues.  A small eigenvalue for $\mathcal{F}'$ will
simply indicate a poorly constrained direction in the $\vec{q}$ space for this
experiment.  For example if $\vec{q}$ is of higher dimension than $\vec{p}$,
then the constraint on $\vec{q}$ must be degenerate at the Fisher-matrix level
of approximation.

If one wants to add a prior on some quantity that is not a single parameter of
the Fisher matrix, one can work in variables where it is a single parameter and
then transform the diagonal $\widehat{\mathcal{F}}^p$  (where the hat indicates
that the new variables are used) back into the working variables using the
inverse of the transformation given in Eq.\ (\ref{steen}).  In
general, if the prior can be expressed as a likelihood over the
parameter set, it can be assigned a Fisher matrix via Eq.\ (\ref{flnp}).

\subsection{Marginalization}
\label{marginalization}

On many occasions we need to produce a Fisher matrix in a smaller parameter
space by marginalizing over the uninteresting ``nuisance'' parameters.  This
amounts to integrating over the nuisance parameters without assuming any
additional priors on their values. 

Suppose the full parameter vector set is $\vec{p}$, which is a union of two
parameter sets: $\vec{p}=\vec{q}\cup\vec{r}$, and we are really
only interested 
in the Fisher matrix for the parameter set $\vec{q}$.  The Fisher matrix
$\cal{F}^\prime$ for parameters $\vec{q}$ after marginalization over
$\vec{r}$ can be expressed as
\begin{equation}
\mathcal{F}^\prime = \mathcal{F}_{qq} 
- \mathcal{F}_{qr} \mathcal{F}_{rr}^{-1} \mathcal{F}_{rq}.
\label{marg1}
\end{equation}
Here $\mathcal{F}_{rr}$, etc., are submatrices of $\mathcal{F}$. 

A common, but numerically unstable procedure, is to 
 
\begin{enumerate} 

\item invert $\mathcal{F}$,

\item remove the rows and columns, corresponding to $\vec{r}$, that are being
marginalized over,

\item then invert the result to obtained the reduced Fisher matrix.

\end{enumerate}

This will fail for an
ill-conditioned $\mathcal{F}$.  Using Eq.\ (\ref{marg1}) is more stable, since
degeneracies in $\mathcal{F}_{qq}$ are properly propagated to
$\mathcal{F}^\prime$ with no instability.  One need worry only about an
ill-conditioned $\mathcal{F}_{rr}$ submatrix, i.e., the experiment has a poor
constraint on some of the marginalized parameters.  One should check for this
by diagonalizing $\mathcal{F}_{rr}=U \Lambda U^T$, where $\Lambda={\rm
diag}(\lambda_1, \lambda_2, \ldots)$ and $U$ is orthogonal.  Now
\begin{equation}
\mathcal{F}^\prime = \mathcal{F}_{qq} 
- \left(\mathcal{F}_{qr}U\right) \Lambda^{-1} \left(\mathcal{F}_{qr}U\right)^T.
\label{marg2}
\end{equation}
If one finds that $\lambda_i$ is small, then one needs to examine row
$i$ of $\mathcal{F}_{qr}U$: if it is (near) zero, then this degeneracy in $r$
does not propagate into parameters $q$ upon marginalization, and it is
safe to ignore this eigenvector.  In other words: if there is some
component to $r$ that is uncorrelated with $q$, we do not care whether
$r$ was constrained by the experiment.

The remaining case where roundoff is a potential problem is when $\Lambda$ has a
small eigenvalue that does not have a correspondingly small row in
$\mathcal{F}_{qr}U$.  This means that there was a component of $\vec{q}$ that
was highly covariant with a component of $\vec{r}$, such that marginalization
over $r$ can leave the $q$ component poorly constrained.  Roundoff error might
lead to $\mathcal{F}^\prime$ indicating a weak constraint on $q^i$ when there
should be none.  But our exposure to this kind of problem should be very
small, since the fallacious constraints induced by roundoff error will
be weak.

When we are combining the constraints from experiments $A$ and $B$, we will be
summing their Fisher matrices. In general any marginalization over nuisance
parameters must be done {\it after} summation of the two Fisher matrices.  If,
however, the nuisance parameters of $A$ are disjoint from those of $B$, then the
two data sets have independent probability distributions over the set of
nuisance parameters, and it is permissible to marginalize before summation.

\section{The Fiducial Cosmological Model}
\label{sec:fiducial}

We have to choose a fiducial cosmological model.  We choose the standard
Friedmann-Lema\^itre-Robertson-Walker cosmology. Our fiducial cosmological model
also assumes that $w(a)=-1$. The other parameters of the fiducial cosmological
model relevant for dark energy data models are (note that the numbering begins
at $0$ and not $1$).

\begin{enumerate}
\setcounter{enumi}{-1}

\item $n_S$, the primordial scalar perturbation spectral index.

\item $\omega_M$, the present value of $\Omega_Mh^2$, where $\Omega_M$ is the
ratio of the present value of the ratio of the total (baryons plus dark matter)
mass density, $\rho_M$, to the critical density, $\rho_C=3H_0^2/8\pi G$, and
$h$ is the reduced Hubble constant, $H_0 = 100h \textrm{  km
s}^{-1}\textrm{Mpc}^{-1}$.

\item $\omega_B$, the present value of $\Omega_Bh^2$, where $\Omega_B$ is the
ratio of the present value of the ratio of the baryon mass density to the
critical density.

\item $\omega_k$, the present value of $\Omega_kh^2$, where $\Omega_k$ is the
ratio of the present value of the ratio of the effective curvature mass density
to the critical density.  From the Friedmann equation,
$\Omega_k=1-\Omega_M-\Omega_{DE}-\Omega_R$, where $\Omega_R$ is the contribution
of radiation.

\item $\omega_{DE}$, the present value of $\Omega_{DE}h^2$, where $\Omega_{DE}$
is the ratio of the present value of the ratio of the dark energy density to the
critical density.

\item $\Delta\gamma$, the change in the growth factor from the standard GR
result.  So the actual growth factor is $f_{\rm actual}(z) = f_{\rm GR}(z)[1 +
\Delta\gamma\ln\Omega_M(z)]$.  This forces $f(z)\rightarrow 1$ at high
redshift. 

\item $\Delta\cal{M}$, the difference between the absolute magnitude of a
type Ia supernova and its fiducial value. 

\item $G_0$, the normalization of the linear growth function in the era when
JDEM has sensitivity to growth, taken to be $z<9$.  It is given relative to the
GR predication of growth between the recombination era and $z=9$: $G_{\rm
actual}(z=9) = G_0\times G_{\rm GR}(z=9)$.  It is assumed that any decaying mode
excited during the Dark Ages is gone by this time (i.e., before we can plausibly
observe it).  The fiducial value is $G_0=1$. 

\item $\ln\Delta_\zeta^2(k_\star)$, where $\zeta$ is the primordial curvature
perturbation and $k_*$ is the pivot scale, here chosen to be $0.05 \textrm{ 
Mpc}^{-1}$ (note: {\it not} $0.05\,h \textrm{  Mpc}^{-1}$).

\end{enumerate}

Some comments are in order.  

\begin{enumerate}

\item We also need to state the definition of the nonlinear power spectrum with
$\Delta\gamma$ turned on.  Since this is not a fundamental parameter but a
phenomenological one there are many possible definitions.  We are using the
definition that at a particular redshift, the $\Delta\gamma$ parameter does {\it
not} change the mapping from the linear-to-nonlinear $P(k)$ (the NL $P(k)$ only
changes through implicit dependence of linear $P(k)$ on $\Delta\gamma$).  This
prevents an experiment from measuring the full $P(k)$ at one redshift and
claiming a constraint on $\Delta\gamma$; one has to measure the evolution in
$P(k)$ or have velocities in order to do this.

\item It is not necessary to specify the fiducial value of $\cal{M}$.

\item We need not specify $H_0$, since  $\omega_M + \omega_{DE} + \omega_k=h^2$, so
$H_0$ is a derived value.

\end{enumerate}

We take the values of the parameters of the fiducial cosmology to be the 
$\Lambda$CDM model of WMAP5:
\begin{eqnarray*}
n_S                   		& = & 0.963  \\
\omega_M              		& = & 0.1326 \\
\omega_B              		& = & 0.0227 \\
\omega_k              		& = & 0      \\
\omega_{DE}           		& = & 0.3844 \\
\Delta\gamma          		& = & 0      \\ 
\Delta\cal{M}         		& = & 0      \\
\ln G_0               	   	& = & 0      \\
\ln\Delta_\zeta^2(k_\star) 	& = & -19.9628   \ \ . 
\end{eqnarray*}

Using $\omega_M + \omega_{DE} + \omega_k=h^2$, the derived value of $h$ is
0.719, which implies a Hubble constant of $H_0=71.9\textrm{ km s}^{-1}
\textrm{Mpc}^{-1}$. Our choices also imply $\sigma_8=0.798$.

\subsection{Modeling the dark-energy equation of state parameter $w$
\label{Modeling}}

While the DETF model for $w(a)$ was very simple: $w(a) = w_0 + (1-a)w_a$, here
we wish to consider a more general behavior.  We will model $w(a)$ as piecewise
constant values of $w(a)$ defined in bins of $\Delta a=0.025$ starting at
$a=0.1$ so that there are 36 bins total.  So we will define $w_i$,
$i=0,\ldots,35$ to be a number describing the value of $w$ in the interval
between $a=1-0.025i$ and $a=1-0.025(i+1)$ or equivalently between
$z=0.025i/(1-0.025i)$ and  $z=0.025(i+1)/[1-0.025(i+1)]$.

In the fiducial model, $w_i=-1$ for $i=0,\ldots 35$.

\subsection{Format of the Fisher Matrix \label{Format}}

Here we establish the ordering and format of the Fisher matrix.  Since Fisher
matrices for different techniques will be combined, the format for every Fisher
matrix must be identical.

When we specify the values of the Fisher matrix, we use the ordering
\begin{eqnarray*}
0 \phantom{XXXXXXX} & & n_S  \\
1 \phantom{XXXXXXX} & & \omega_M              \\
2 \phantom{XXXXXXX} & & \omega_B              \\
3 \phantom{XXXXXXX} & & \omega_k              \\
4 \phantom{XXXXXXX} & & \omega_{DE}           \\
5 \phantom{XXXXXXX} & & \Delta\gamma          \\ 
6 \phantom{XXXXXXX} & & \Delta\cal{M}         \\
7 \phantom{XXXXXXX} & & \ln G_0               \\
8 \phantom{XXXXXXX} & & \ln \Delta_\zeta(k_*) \\
9 \phantom{XXXXXXX} & & w_0		    \\
10 \phantom{XXXXXXX} & & w_1		    \\
\vdots \phantom{XXXXXXX} & & \vdots		    \\
44 \phantom{XXXXXXX} & & w_{35}
\end{eqnarray*}
Note that the ordering begins with $0$, not $1$, and for $0\leq i \leq 35$,
entry $9+i$ corresponds  to $w_i$.

We denote by $\mathcal{F}_{ij}$ the $ij$ component of the Fisher matrix.  The
Fisher matrix is symmetric.  By convention, unless specified, it is assumed that
$\mathcal{F}_{ij}=0$. The Fisher matrix is supplied in ascii format in the form
\begin{equation}
i \ \ \ \ j \ \ \ \ \mathcal{F}_{ij}.
\end{equation}
A line that begins with a ``\#'' is considered to be a comment line.

It is necessary to specify the Fisher matrix in double precision.

\section{Dark Energy Circa 2016}
\label{sec:circa}

Here we will not go into detail about how the pre-JDEM\footnote{In this section
we discuss projections of knowledge of dark energy parameters in the
``pre-JDEM'' time frame.  To be explict, by pre-JDEM projects we mean projects
that we expect will have a cosmological likelihood function available before
JDEM is begins its first data analysis.} Fisher matrices are constructed,
rather, we relegate that to an appendix.  However, we would like to make the
important point that although considerable effort went into constructing the
Fisher matrices, they should be used with caution, since any data model
purporting to represent the state of knowledge eight years in the future is
highly suspect. 

We note that our assumed improvements in SN measurements for pre-JDEM 
are conservative, in the sense no significant advances in
techniques or astrophysical knowledge are assumed before JDEM. 

For our WL Fisher matrix, we assumed some improvements in the systematic 
error budgets as detailed in the Appendix.  This is reasonable as there 
are identifiable (though not necessarily guaranteed) paths toward 
improving these errors.  We have also been conservative in not assuming 
the use of all WL information (e.g., we do not include the bispectrum, and 
included only one of the WL surveys) or any theoretical knowledge about 
intrinsic alignments.  Overall we believe the WL Fisher matrix is somewhat 
conservative but not extremely conservative.

We have been more optimistic about the expected advances in our knowledge for
the BAO technique, e.g., measuring the BAO scale at high redshift using
Lyman-alpha galaxies or absorbers, and the reconstruction method of Eisenstein
et al.\ \cite{Eisenstein2007}. We feel this is justified as the strength of the
BAO method is that the distance between optimism and pessimism is smaller (see
the DETF report).

\subsection{Prospective Additional Probes of Dark Energy}

\subsubsection{Galaxy Clusters (Number Density, Clustering and Their Evolution)}

The abundance and clustering of galaxy clusters is another promising technique,
and has previously been considered by the DETF \cite{DETF}.  There are many
means of identifying and measuring galaxy clusters; the main source of
uncertainty in future applications of this method will be in determining the
relation of the selection function and observables to the underlying mass of the
clusters.  In this report we chose not to include clusters in our pre-JDEM
Fisher matrix,  because we were unable to forecast its level of systematic error
in mass determination ca.\ 2016 with any useful degree of reliability. Since all
methods are challenged by sources of systematic errors, some common among them
and some independent, there is value in addressing dark energy with all feasible
techniques, including cluster studies.  Here, we briefly review the potential of
the cluster technique.

Clusters are a very promising technique, in particular because X-ray,
Sunyaev-Zeldovich (SZ) and optical and near-infrared (NIR) data can be combined
to minimize systematic errors in identifying and characterizing the cluster
population.  Present cluster data and systematic errors determine cosmological
parameters, particularly $\sigma_8$, to accuracy similar to present weak-lensing
data.

Mass selection systematics are smallest for X-ray- or SZ-selection
\cite{Rosati}, and there is concrete hope for further reduction in these
uncertainties in the near future. Within the halo model framework, key elements
for cluster cosmology are the form of the observable signal likelihood as a
function of mass and redshift, and the effects of projection (including
non-cluster sources) on measured signals. X-ray-selected surveys have
particular advantage in the latter aspect; although AGN emission is a
contaminant that can only be removed with sufficiently high-resolution X-ray
data. All signals require empirical calibration of low-order moments,
particularly variance, in the scaling relations.  X-ray \cite{Rykoff,Pratt} and
SZ \cite{Kravtsov,Arnaud} signals have local calibrations with statistical and
systematic errors  in slope and intercept at the level of about 10\%, with
considerable potential to improve with higher quality data.

Compared to X-ray or SZ methods, optical- and NIR-selected cluster surveys are
more strongly affected by blending of multiple halos in projection, a
long-standing problem known to Abell. Multi-color information provides the
photometric redshift information that can significantly reduce projection
effects.  Estimates for single-color photometric selection indicate that
low-mass leakage (i.e., smaller halos that, viewed in alignment, appear as a
single, rich system) adds a 10 to 20\% enhancement in number relative to the
traditional mass selection \cite{Cohn}.

Optical samples offer the benefit of extending the cluster population to lower
masses than is achievable by X-ray and SZ. This could potentially decrease the
shot noise by factors of several if confusion due to projection can be
accurately modeled in the low richness regime. Theoretical elements needed to
achieve successful modeling of projection effects are: 1) the halo mass
function, 2) halo clustering bias, and 3) multi-color galaxy occupation models
that describe both central and satellite galaxies in halos out to redshifts
$z\sim 1.5$. The form of this  last element, which is only now being studied,
is a crucial source of systematic uncertainty that must be overcome in order
for optical richness to function as a viable mass proxy.

The situation will improve significantly when deep samples dramatically
increase the current sparse numbers of high clusters, allowing
``self-calibration'' of scaling relations \cite{MM2004,Lima_Hu}. In imagining
future developments, one should keep in mind that all current measurements of
cosmological parameters with clusters are based on samples of only a few
hundred objects \cite{Rosati,Schuecker,Allen,Vikhlinin}.  In light of this,
one should be optimistic given the set of funded projects that will return
data in the next decade: 1) South Pole Telescope (SPT) and Atacama Cosmology
Telescope (ACT) will reveal many thousands of new SZ clusters above about 
$2\times 10^{14} M_\odot$ over several thousand sq deg; 2) Dark Energy Survey
(DES) and UK VISTA will overlap the SPT area with optical/NIR for photo-$z$
estimates and additional mass calibration, and 3) a new X-ray survey mission,
e-Rosita, will launch in 2012 and cover $20,000$ sq.\ deg.\ to a flux limit
nearly two orders of magnitude fainter than the present ROSAT All-Sky Survey.
Combining homogeneous X-ray, SZ and optical/NIR samples across large areas of
sky will provide unprecedented sensitivity to distant clusters.  Such data,
extended with targeted, multi-wavelength follow-up by ground- and space-based
observatories, will provide critical cross-calibration abilities that will
serve to mitigate sources of systematic error associated with cluster
selection and signal--mass characterization.

Galaxy clusters can also be detected and measured using weak gravitational
lensing shear.  Since lensing shear responds directly to mass, systematic
errors in mass assignment are bypassed.  This has been exploited to determine
the mean mass of x-ray or optically selected clusters via ``stacking'' of
lensing signals \cite{Johnston:2007uc}.  Future deep WL surveys could usefully
detect clusters, or, more precisely, shear peaks.  Any JDEM executing WL shear
tomography would obtain the necessary data for this cosmological test.  Such a
technique removes the power spectrum from the shear tomography measurement,
resulting in a purely geometric determination of angular distance ratios. 
While such a measurement will have a lower signal-to-noise ratio than the full
WL survey, it would provide an internal check for any systematic errors in the
survey.

\subsubsection{Redshift-Space Distortions}

Redshift-space distortions in the clustering pattern of galaxies have recently
been the subject of renewed significant attention in the context of dark energy
as a method to trace the evolution of the growth rate of structure with redshift
\cite{Guzzo08}.  The linear part of redshift distortions is induced by the
coherent motions of galaxies partaking in the overall growth of overdensities. 
By measuring galaxy two-point correlations along and perpendicular to the
line-of-sight, one can quantify the distortions in terms of the linear
compression  parameter $\beta$, which is related to the growth rate $f$ as
$f=b_L\times \beta$ \cite{Kaiser87} (in this expression $b_L$ is the {\it linear
bias} parameter of the class of galaxies being used as tracers of large-scale
structure).  This provides, in principle, a way to detect evidence for modified
gravity as an alternative solution to the accelerated expansion problem. 
Conversely, if General Relativity is assumed, this technique provides one
further method to constrain the evolution of the equation of state $w(a)$,
through a precise measurement of $\Omega_M(z)$ at different
redshifts\cite{Wang08, Linder08}.

The practical application of redshift-space distortions as an acceleration probe
is  still in its infancy: spectroscopic samples beyond $z>0.5$ are just now
starting to be large enough as to allow extraction of the linear signal from the
noise, overcoming at the same time cosmic variance. A positive aspect, however,
is that due to the intense investigations performed during the 1990's on local
samples (as e.g., the IRAS surveys), there exist significant technical knowledge
on how to account and correct for systematic and nonlinear effects
\cite{Hamilton98}.  Monte Carlo forecasts based on mock surveys indicate that it
is possible to reach statistical errors on $\beta$ of the order of a few percent
within   surveys extending to $z\sim 1$ with densities of objects of order
$10^{-2}$ to $10^{-3}h^3$ Mpc$^{-3}$ and covering volumes about $10^8$ to
$10^9h^{-3}$ Mpc$^3$ \cite{Guzzo08}.  A JDEM-class survey providing 100
million redshifts spread through 10 redshift bins between 0.5 and 2, would reach
similar  percent accuracy on $\beta$ within each bin, constraining not only a
simple ``$\gamma$-model'' for the growth rate, but also the detailed evolution
of $f(z)$ itself \cite{Aquaviva08}.   This is for example the kind of survey
planned by the European mission EUCLID.

In fact \cite{SongPercival08,PercivalWhite08}, redshift-space distortions
constrain the combination $\sigma_8(z)\times f(z)$, which in the assumption of
linear deterministic bias, corresponds to the observed product $\beta \times
\sigma_8(\textrm{galaxies})$.   When combined with the high-redshift
normalization of the power spectrum  from CMB observations (e.g., Planck), this
enables one to determine the  product $G(z)f(z)$.  An independent determination
of the galaxy bias, e.g., from the bispectrum \cite{2002MNRAS.335..432V}, would
enable both $G(z)$  and $f(z)$ to be measured separately.  Recent theoretical
work \cite{McDonald_Seljak08, White08} suggests in addition that using two
classes of galaxy tracers with different bias factors, should allow one to
reach even higher accuracies.  This awaits confirmation through mock surveys
with realistic noise properties, however, it is a  clear indication of the
active developments in this area.  While current constraints on the growth rate
are at the several tens of percent level (e.g., \cite{Guzzo08,SongPercival08}),
progress is rapid. It is reasonable, therefore, to predict redshift-space
distortions to emerge in the next decade as one of the main tools of
investigation for cosmic acceleration. 

\subsubsection{The Integrated Sachs-Wolfe (ISW) Effect}

When a photon passes through a time-dependent gravitational potential, its
energy changes.  This effect produces a secondary (i.e., not primordial)
anisotropy in the CMB \cite{1967ApJ...147...73S} that is correlated with large
scale structure \cite{1996PhRvL..76..575C}.  This anisotropy is known as the
integrated Sachs-Wolfe (ISW) effect.  Since gravitational potentials vary with
cosmic time according to $\Phi\propto G(a)/a$, a cross-correlation of the CMB
with large-scale structure at a particular redshift $z$ allows one to probe the
rate of growth of cosmic structure at that redshift.  The signal depends on the
rate of change of gravitational potential $\dot\Phi$, which is proportional to
$(f-1)G$.  This is very attractive from a systematics point of view since
$f\approx 1$ during most of the history of the universe:\footnote{We have $f=1$
in the matter-dominated era, assuming GR.} Nature takes the difference $f-1$
for us, unlike redshift-space distortions (proportional to $Gf$) or weak
lensing (where $G(z)$ is measured using source galaxies at different redshifts,
and then differentiated to get $f$).  Recent ISW analyses combining WMAP with
low-redshift data have come close to their ideal statistical error limits
\cite{2008PhRvD..78d3519H, 2008PhRvD..77l3520G} and the technique has one of
the highest levels of maturity.  The largest systematic has been chance
correlations of Galactic foregrounds on the largest angular scales with
extragalactic structures; microwave emission from extragalactic objects (even
AGNs) has turned out to be subdominant to the ISW effect at the angular scales
of interest.  With the broad frequency coverage of the upcoming Planck mission
it is reasonable to expect that in the 2016 timeframe the ISW effect will be
essentially statistics-limited outside the Galactic Plane region.

The downside to the ISW effect is that the ISW signal has exactly the same
frequency dependence as the primordial CMB anisotropies, and hence the latter
constitute an irreducible noise source.  Indeed, it is expected that the latter
will limit the ISW detection significance approximately to the $7.5 f_{\rm
sky}^{1/2}\;\sigma$ level \cite{2004PhRvD..70h3536A}. Constructing a Fisher
matrix based on the signal and noise estimates of Afshordi
\cite{2004PhRvD..70h3536A} we find that even a ``perfect'' ISW program out to
$z=2$ covering $2/3$ of the sky and reaching the sampling variance limit ($nP\gg
1$) makes little improvement to the dark energy constraints: for example, adding
it to our pre-JDEM prior improves the DETF FoM from 116 to 120.  It would have
an effect on modified gravity constraints: $\sigma(\Delta\gamma)$ improves from
0.21 to 0.08.  We emphasize that this is {\em not} included in the FoMSWG
pre-JDEM standard prior, since not all of the relevant large-scale structure
data will exist in 2016.  We also note that ongoing studies of JDEM weak-lensing
programs suggest that these can constrain $\Delta\gamma$ much more tightly than
even the perfect ISW program.  That said, the ISW effect will always be valuable
as a cross-check, and as a test of more general modified gravity phenomenologies
(e.g., scale-dependent $\Delta\gamma$).  It may also provide a competitive
constraint on $\Delta\gamma$ in the pre-JDEM era, depending on the level of
systematics and nuisance parameter requirements encountered by other probes of
the growth of structure.

\subsubsection{The Alcock-Paczynski Effect  \label{sec:AP}}

The anisotropic clustering of galaxies also contains a probe of the distance
scale through the Alcock-Paczynski effect \cite{AP79, Ballinger96, 
Matsubara02, Shoji08}.  This provides a measurement of $H(z)\times D_A(z)$,
which is important by itself but becomes particularly useful as a way to
convert constraints on $D_A(z)$ from BAO or SNe into constraints on $H(z)$. 
However, this effect is mixed together with the dynamical redshift-space 
distortions and their interaction with galaxy bias.  The relative  weight of
these two effects depends on the redshift and their signatures are in principle
separable through detailed modeling.   At $z\sim 1$ dynamical distortions
dominate \cite{Guzzo08} and attempts to model both from galaxy data show
strong degeneracy \cite{Ross}.  Similarly to the case of dynamical redshift
distortions, improved theoretical modelling  in the coming decade could change
this situation and unlock a significant  new test of dark energy from galaxy
redshift surveys.

\section{On quantifying the merit of dark-energy missions \label{quant}}
\label{sec:foms}

In this section we describe how to extract  useful figure(s) of merit from the
$45\times 45$ Fisher matrix describing any suite of experiments, which
succinctly summarize progress toward the goals described in Sec.\ 
\ref{sec:task}.

Each Fisher matrix is of dimension $45\times 45$, corresponding to nine
cosmological parameters and 36 piecewise constant equation of state parameters
$w_i$, $i\in \{0, 35\}$.  If a method is insensitive to a given parameter $p^k$,
then the corresponding Fisher elements $\mathcal{F}_{ik}$ will be zero for each
$i$, but we still insist on retaining the full dimensionality of the matrix. 
Each technique $t$ has a Fisher matrix $\mathcal{F}^t$. We can consider
combinations of techniques by adding the corresponding Fisher matrices.  The
total efficacy can be found by summing the Fisher matrices for all techniques:
$\mathcal{F}^\textrm{total}=\sum_t \mathcal{F}^t$. We suggest that the
procedures we outline herein should be performed using the Fisher matrix for
each JDEM technique individually, as well as the techniques in combination.

\subsection{Tests of smooth DE and \LCDM}

\subsubsection{Principal components}

In the principal component approach \cite{Hut_Stark}, we aim to understand the
redshifts at which each experiment has the power to constrain $w$.  One starts
by expanding the equation of state as
\begin{equation}
1+w(a) = \sum_{i=0}^{35} \alpha_i e_i(a),
\label{eq:PC_expansion}
\end{equation}
where $\alpha_i$ are coefficients, and $e_i(a)$ are the eigenvectors to be
defined below.  The procedure for computing the principal components goes as
follows:

\begin{enumerate} 

\item Since in this section we are considering constraints on the equation of
state, we fix the parameters $\Delta\gamma$ and $G_0$ to their fiducial values
($0$ and $1$, respectively). This can be done by adding a strong prior to the
corresponding diagonal elements of $\mathcal{F}$ ($\mathcal{F}_{55}$ and
$\mathcal{F}_{77}$), as described in Sec.\ \ref{sec:addprior}.

\item Now we wish to marginalize over all cosmological parameters except those
corresponding to $w_i$. Section \ref{marginalization} describes the matrix
manipulations that implement marginalization.

\item Now rotate the parameters into a new basis by diagonalizing
the marginalized Fisher matrix 
\begin{equation}
\mathcal{G} = \mathcal{W}^T {\bf \Lambda} \mathcal{W},
\label{eq:diagonalize}
\end{equation}
where ${\bf \Lambda}$ is diagonal and $\mathcal{W}$ is an orthogonal matrix.
It is clear that the new parameters $e_i$, defined as ${\bf e}\equiv
\mathcal{W}{\bf p}$, are uncorrelated, for they have the diagonal covariance
matrix  ${\bf \Lambda}^{-1}$. The $e_i$ are referred to as the principal
components and the rows of $\mathcal{W}$ are the window functions (or weights)
that define how the principal components are related to the $p_i$. For a fixed
$i$, the $i^{th}$ row of $\mathcal{W}$ is the eigenvector $e_i(a_m)$, where
$a_m$ runs over the (discrete) values of the scale factor; $a_m\equiv
1-(m+1/2)\Delta a$ with $m\in \{0, 35\}$. Elements of the matrix ${\bf  
\Lambda}$ are the corresponding  eigenvalues $\lambda_i$.

\item We normalize the amplitude of each eigenvector so that 
\begin{equation}
\int e_i^2(a)da\rightarrow  \sum_m e_i^2(a_m)\Delta a=1,
\label{eq:normalize}
\end{equation}
(where, recall, $\Delta a=0.025$ is the binning width in scale factor), and we
correspondingly renormalize the eigenvalues $\lambda_i$ as well. We can now
compute the coefficients of the expansion of $1+w(z)$ in
Eq.\ (\ref{eq:PC_expansion}) as
\begin{equation}
\alpha_i = \int [1+w(a)] e_i(a)da \rightarrow  
\sum_m [1+w(a_m)] e_i(a_m)\, \Delta a.
\end{equation}

We do not explicitly use the coefficients $\alpha_i$ (note that their values
are all trivially zero for a $\Lambda$CDM model anyway). We are however
interested in the {\it variance} in these coefficients:
\begin{equation}
\sigma^2_i \equiv \sigma^2(\alpha_i) = \frac{1}{\lambda_i}.
\end{equation}
Finally, we rank-order the eigenvectors from best to worst measured, that is,
by the value of $\sigma^2_i$ (smaller $\sigma^2_i$ have higher rank).

\end{enumerate}

We choose the variance, $\sigma_i^2$, to be the appropriate measure rather than
$\sigma_i$.  Measurements that are limited by statistical errors have their
variance improve inversely with the size of the data set.  Hence, we focus on
the ratio of variances as this can be interpretted as the ratio of the survey
size, e.g., as measured by cost, effort, or time.\footnote{Of course,
measurements that are limited by systematic errors or by parameter  degeneracies
broken only by external data sets will improve more slowly with size; it is not
necessarily the case that one can improve the variance by a factor of 10 simply
by running the experiment 10 times  longer.}

A principal component is considered usefully constrained only if its variance
$\sigma^2_i$ is less than unity, since in that case it limits the RMS
fluctuation of the equation of state in that mode, $\langle
[1+w(z)]^2\rangle=\sigma_i^2$, to be smaller than unity.  In other words we
have a prior prejudice that variations in $w$ beyond $O(1)$ are unlikely. We
impose this prior belief by performing the following operation
\begin{equation}
\frac{1}{\sigma_i^2} \rightarrow 1 + \frac{1}{\sigma_i^2} \quad \Rightarrow 
\quad \sigma^2_i \rightarrow \frac{\sigma^2_i}{1+\sigma_i^2},
\label{eq:PCprior}
\end{equation}
This procedure regularizes the PC errors, i.e., sets $\sigma^2(\alpha_i)$ to be
unity at worst, making the quantitative nature of the PC error plots (described
below) easier to understand. For our pre-JDEM data, 15--20 of the
best-determined PCs have raw variance $\sigma^2_i<1$, and approximately
the first $10$ of those are negligibly affected by the prior in Eq.\
(\ref{eq:PCprior}). Since the JDEM principal components will be determined even
better than the pre-JDEM ones, we see that our prior does not significantly
affect the first 10--20 components, allowing the reader to infer the true
intrinsic power of JDEM to probe $w(z)$.

The remaining step is to plot the principal components and the accuracies with
which they are measured. We find that two plots are particularly useful:

\begin{enumerate}

\item A plot of the first three (or so) eigenvectors $e_i(z)$. We choose
  plotting them in redshift $z$ rather than scale factor $a$ since the former is
  more commonly used in observational cosmology.  The best-measured eigenvectors
  most clearly indicate the redshifts at which the probe is capable of
  detecting variations in dark-energy density.

 \item A plot of the eigenvector variance, $\sigma^2_i$ vs.\ $i$ for
   the first 10 to 20 eigenvectors.  It is most useful to quantify how much
   JDEM improves upon the pre-JDEM errors.  This can be done by plotting the
   ratio 
   \begin{equation}
     \frac{\sigma^{-2}_i({\rm JDEM)}}{\sigma^{-2}_i(\textrm{pre-JDEM})},
   \end{equation}
   where pre-JDEM errors (that we project and provide in the appendix)
   approximate the accuracies available from combined ground-based surveys in
   year 2015.  This ratio is generally greater than unity for any PC and,
   because of the regularization procedure in Eq.\ (\ref{eq:PCprior}), the
   ratio will approach unity for poorly measured modes.  The idea is to plot
   all accuracies that are measured better than unity, that is, up to the
   highest value of $i$ for which $\sigma^2_i\simeq 1$ (at which point the prior
   from Eq.\ (\ref{eq:PCprior}) completely dominates $\sigma^2_i(\rm JDEM)$).

\end{enumerate}

\begin{figure}[t]
\centering\leavevmode
\includegraphics[width=4.5in,angle=0]{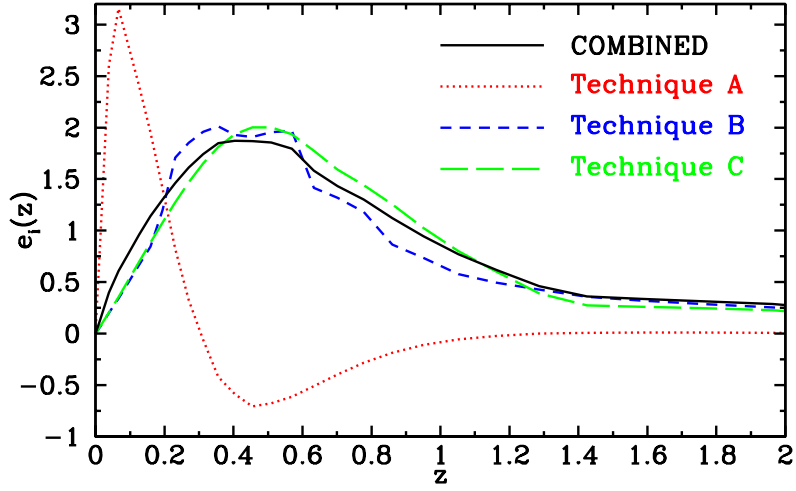}
\caption{An example of the first principal component for individual techniques, 
and all of the techniques combined.}
\label{PC}
\end{figure}

An example of a graph of the first principal components of a JDEM mission is
shown in Fig.\ \ref{PC}.  In this example we see that technique A is
sensitive to lower redshifts than the other techniques.  The redshift
sensitivity of the combined probes is dominated by techniques B and C in this
particular example, because the two have very similar best-measured
eigenvectors as the combination of the three.  

\begin{figure}[t]
\centering\leavevmode
\includegraphics[width=4.5in,angle=0]{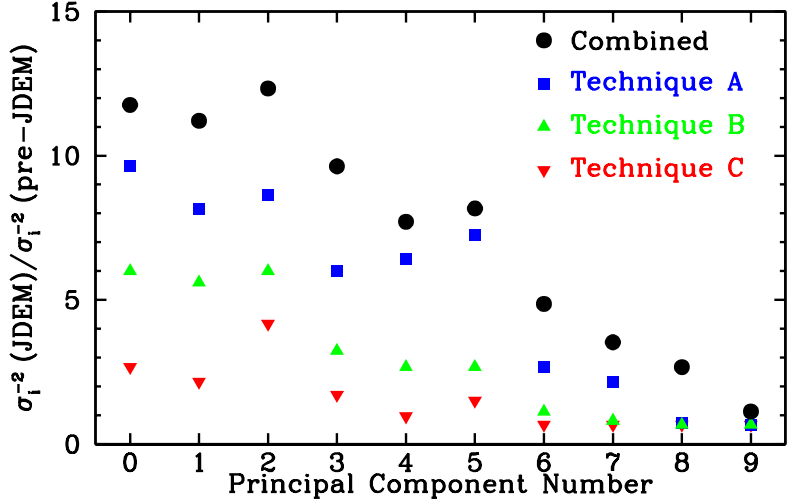}
\caption{An example of how one might present a graph of the increase in 
knowledge of the evolution of dark energy from a JDEM mission, normalized to the
pre-JDEM knowledge. The data points shown in this figure do not correspond to 
any proposed JDEM, but invented merely to illustrate an example graph.}
\label{S}
\end{figure}

Figure \ref{S} shows the corresponding ratios $\sigma^{-2}_i({\rm JDEM})/
\sigma^{-2}_i(\textrm{pre-JDEM})$.  To construct this figure, it is necessary to
know the values of $\sigma_i(\textrm{pre-JDEM})$.  They are given in Table
\ref{tab:sigma}.

\begin{table}
\begin{ruledtabular}
\begin{tabular}{rcccc}
 i  & $\sigma_i(\textrm{BAO})$ & $\sigma_i(\textrm{SN})$ & 
 $\sigma_i(\textrm{WL})$ & $\sigma_i(\textrm{COMBINED})$ \\ \hline \hline
 1  &     0.027       &      0.103     &      0.061     &   0.021  \\
 2  &     0.058       &      0.211     &      0.134     &   0.044  \\
 3  &     0.090       &      0.381     &      0.231     &   0.069  \\
 4  &     0.116       &      0.567     &      0.337     &   0.094  \\
 5  &     0.158       &      0.768     &      0.493     &   0.121  \\
 6  &     0.190       &      0.883     &      0.628     &   0.158  \\
 7  &     0.250       &      0.956     &      0.657     &   0.188  \\
 8  &     0.289       &      0.984     &      0.783     &   0.239  \\
 9  &     0.348       &      0.993     &      0.851     &   0.276  \\
10  &     0.396       &      0.998     &      0.922     &   0.324  \\
11  &     0.418       &      0.999     &      0.943     &   0.386  \\
12  &     0.456       &      1.000     &      0.980     &   0.390  \\
13  &     0.533       &      1.000     &      0.996     &   0.451  \\
14  &     0.581       &      1.000     &      0.998     &   0.521  \\
15  &     0.620       &      1.000     &      0.999     &   0.534  \\
16  &     0.662       &      1.000     &      0.999     &   0.595  \\
17  &     0.734       &      1.000     &      1.000     &   0.668  \\
18  &     0.768       &      1.000     &      1.000     &   0.726  \\
19  &     0.801       &      1.000     &      1.000     &   0.764  \\
20  &     0.851       &      1.000     &      1.000     &   0.795  \\
21  &     0.900       &      1.000     &      1.000     &   0.824  \\
22  &     0.926       &      1.000     &      1.000     &   0.862  \\
23  &     0.945       &      1.000     &      1.000     &   0.905  \\
24  &     0.961       &      1.000     &      1.000     &   0.929  \\
25  &     0.970       &      1.000     &      1.000     &   0.955  \\
\end{tabular}
\end{ruledtabular}
\caption{Pre-JDEM values of $\sigma_i$ for the first 25 best determined
principal components for various techniques, plus the techniques combined.
\label{tab:sigma}}
\end{table}

\subsubsection{$(w_0, w_a)$ and the pivot values}

The DETF FoM remains a useful simple measure of an experiment's constraining
power.  Since the 36-parameter description of the equation of state is quite
general, it is clear that we should be able to project straightforwardly the
constraints on the $w_i$ onto the parameters $(w_0, w_a)$ used by the DETF. 
Projection of Fisher matrices onto new variables is discussed in Sec.
\ref{sec:projection}. In this part of the analysis, one should {\it not} add
priors to the principal components as in Eq.\ (\ref{eq:PCprior}).

Let $\mathcal{H}$  be the desired $2\times 2$
Fisher matrix in parameters $w_0$ and $w_a$. Then
\begin{equation}
\mathcal{H}_{ij} = \sum_{m, n} 
{\partial w_m\over \partial p^i} \mathcal{G}_{mn} 
{\partial w_n\over \partial p^j},
\end{equation}
where the parameters $p$ are either $w_0$ or $w_a$. Moreover, the derivatives
are trivially simple: $\partial w_m/\partial w_0=1$ and $\partial
w_m/\partial w_a=1-a_m$; here $a_m$ is the scale factor corresponding to
the index $m$. Now that we have the matrix $\mathcal{H}$, the errors in $w_0$
and $w_a$ are easy to obtain: 
\begin{eqnarray}
\sigma^2(w_0) &=&\left(\mathcal{H}^{-1}\right)_{w_0w_0},\\
\sigma^2(w_a)&=&\left(\mathcal{H}^{-1}\right)_{w_aw_a}.
\end{eqnarray}

Recall that the goal is to test whether or not dark energy arises from a
cosmological constant ($w_0=-1$, $w_a=0$ in the DETF parametrization).  For
each data model the constraint on $w(a)$ varies with $a$.  However, there is
some pivot value of $a$, denoted by $a_p$, where the uncertainty in $w(a)$ in
minimized for a given data model.  The value of the scale factor (or redshift)
with the minimun uncertainty is called the pivot scale factor (or redshift),
$a_p$ (or $z_p$).  One can also define a pivot value of $w$ as
\begin{equation}
\label{wp}
w_p = w _0 + (1-a_p)w_a.
\end{equation}

Finally, we would also like to quote the pivot redshift $z_p$ and the
uncertainties in $w_p\equiv w(z_p)$. Since $w_p$ is by definition uncorrelated
with the parameter $w_a$, it is easy to show that
\begin{eqnarray}
 z_p &=& -\frac{\left(\mathcal{H}^{-1}\right)_{w_0w_a}} 
               {\left(\mathcal{H}^{-1}\right)_{w_0w_a} + 
    \left(\mathcal{H}^{-1}\right)_{w_aw_a}}, \\
\sigma^2(w_p)  &=& \left(\mathcal{H}^{-1}\right)_{w_0w_0} - 
\frac{\left[\left(\mathcal{H}^{-1}\right)_{w_0w_a}\right]^2}
     {\left(\mathcal{H}^{-1}\right)_{w_aw_a}}  .
\end{eqnarray}
The DETF FoM is $|\mathcal{H}|^{1/2} = 1/\sigma(w_p)\sigma(w_a)$.

\subsection{Tests of general relativity through growth of structure}

We also report accuracy in the ``growth index'' $\gamma$, defined
in the fitting function for the linear growth of density perturbations:
\begin{equation}
G(a)= G_0\exp\left ({\int_{0.1}^a d\ln a\,[\Omega_M^\gamma(a)-1]}\right ),
\label{eq:groexp}
\end{equation}
where $G(a)\equiv \delta/a$ is the growth of density perturbations normalized to
Einstein-de Sitter case.  The $\gamma$ parametrization with $\gamma=0.55$ was
shown to closely approximate the exact solution within GR for a wide variety of
physical dark energy equation of state ratios \citep{Linder_2005}; measuring a
departure of $\gamma$ from its fiducial value ($0.55$ in $\Lambda$CDM) therefore
provides a test of general relativity. When computing $\gamma$, we allow the
freedom in the parameter $\ln G_0$, and fully marginalize over all other
parameters as well, including the equation of state parameters $w_i$. 

We have found that the highest-redshift $w_i$ bins are very poorly constrained
by the data.  These $w_i$s can be partially degenerate with $\Delta\gamma$
because extremely large fluctuations in the dark energy at early times can
re-excite the decaying mode of linear perturbations, thereby changing $f$ and
affecting the measurement of $\Delta\gamma$. Therefore, when marginalizing over
the $w_i$ (and only when marginalizing over the $w_i$), we add a prior of
$\Delta a/100$ to the diagonal element of the Fisher matrix associated with each
$w_i$.  This prior is extremely weak--it corresponds to penalizing models where
$|1+w|$ exceeds approximately $10$--and it prevents numerical problems
associated with the unconstrained $w_i$s from leaking into
$\sigma(\Delta\gamma)$.

\section{Interpreting the figures and numbers}
\label{sec:interp}

Before presenting some broad guidelines for interpreting the information
generated following the FoMSWG procedure outlined in the above section, it is
important to stress that the FoMSWG feels that it would be ill advised to
pursue a dark energy program on the sole basis of a
figure of merit, or even several such numbers, or even the information
generated following the FoMSWG procedure. Figures of merit, or the result of
the FoMSWG procedure, are only as good as the data models used to generate
them.  Constructing a data model requires assumption about many things, in
particular systematic errors and performance of instruments.  

We feel it is important to emphasize several points:

\begin{enumerate}

\item We are operating under the assumption that the nominal launch date for
JDEM will be sometime in 2016.  Dark energy remains a compelling astrophysical
question (perhaps the {\it most} compelling) and the creativity and imagination
of astronomers and physicists will continue to be directed toward
investigations into the nature of dark energy.  Predictions of what will be
known about dark energy, or what will be known about systematic uncertainties
associated with dark energy measurements, eight years in the future are
inherently unreliable.  What we have reported about this subject is the
informed judgment of the Working Group, but we must emphasize both the
importance, and the inherent uncertainty, of such predictions.    

\item We find that it is impossible to evaluate
different dark energy investigations without the adoption of the same fiducial
cosmological model and prior assumptions as a starting point. 
In this report we proposed a fiducial cosmological model
and appropriate prior information to be used as a starting point.

\item Even with uniform fiducial cosmology and priors, 10--20\% changes in these
figures of merit can result from minor variation in assumptions about
astrophysical parameters in each method, e.g., galaxy biases, SN rates, etc.  

\item The FoMSWG finds that there is no single number that can describe
the scientific reach of a JDEM.  Our efforts have been directed toward
producing a simple prescription that can be used to judge the potential
scientific reach of a given dark energy investigation.  
While again, there is no single number, a few
graphs and numbers can convey a lot of quantitative information.  We have been
prescriptive about this point.

\item Finally, although this is a Figure of Merit Working Group, we feel that it
is important to emphasize that there are important criteria in the evaluation of
an experimental framework or investigation
that are not describable in terms of a figure of merit.  For example,
systematic errors will determine what an investigation will accomplish.  
Allowance for
systematic errors should be part of the procedure used in generating each
investigation's Fisher matrix. Important systematic errors are, however, 
sometimes difficult
to quantify and sometimes even difficult to list.   

\end{enumerate}

With the above {\it caveat emptor}, we do feel that if one starts with reliable
data models the information generated by following the FoMSWG procedure
outlined in Sec.\ \ref {quant} can convey a lot of information.  We now provide
some guidance for interpretation of the various produces.

\begin{enumerate}

\item Graphs of Principal Components as a function of $z$: These graphs provide
information about the redshift coverage and sensitivity of a particular
technique or combinations of techniques.  Because we   have very little
information about the behavior of dark energy with redshift,   it is desirable
to have wide redshift coverage. The sensitivity, on the other hand, is greatest
at the redshift where   the {\it strongest} principal components peak.  The peak
redshift and PC shapes are primarily determined by the cosmological technique
(increasing from   SN to BAO to WL), with a slightly weaker but important
dependence on the   survey specifications. 

\item Graph of $\sigma_i^{-2}(\textrm{JDEM})/\sigma_i^{-2}(\textrm{pre-JDEM})$:
This graph is an indication of the advance of accuracy as the result of a JDEM. 
Recall that the DETF parametrized the evolution of dark energy by
$w(a)=w_0+w_a(1-a)=w_p+w_a(a_p-a)$.  In this parametrization there were two
parameters, which could be chosen to be either $(w_0,w_a)$ or $(w_a,w_p)$ (see
Sec.\ \ref{quant} for a description).  The DETF figure of merit was proportional
to the inverse of the product of the uncertainties in these parameters:
$\textrm{DETF FoM} \propto \left[\sigma(w_a)\times\sigma(w_p)\right]^{-1}$.  Our
corresponding parametrization is much more ambitious, and consists of 36
principal components of the function $w(a)$.  We discussed in Sec.\ \ref{quant}
a regularization procedure that imposes $\sigma^2_i\rightarrow 1$ for large
$\sigma^2_i$, so only usefully determined principal components will   have large
ratios of $\sigma_i^{-2}(\textrm{JDEM})/\sigma_i^{-2}(\textrm{pre-JDEM})$.
Existing computations of principal components show that improving an experiment
tends to improve the measurement of all the independently measured modes in
unison.  Individuals may have different preferences whether to focus on the
first few best-measured modes or a wider range of modes when assessing a given
experiment.  In either case, the graph of
$\sigma_i^{-2}(\textrm{JDEM})/\sigma_i^{-2}(\textrm{pre-JDEM})$ allows one to
make an individual judgment and also note any interesting deviations from the
``improvement in unison'' result that may come up.

\item $(w_0,w_a)$ and the pivot values $(w_p,z_p)$: These numbers are
trivially   generated and provide a useful consistency check of the principal
component   analysis. For instance the pivot value $z_p$ is the redshift where
$w$ is best determined.  This pivot redshift is usually consistent with the  
redshift where the first principal component peaks.  Furthermore, from $w_0$  
and $w_p$ once can construct the DETF figure of merit.  One expects that the
data models that result in better DETF figure of merits will result in larger
values of $\sigma_i^{-2}(\textrm{JDEM})/\sigma_i^{-2}(\textrm{pre-JDEM})$.  Wild
discrepancies would indicate the need to understand the reason.

\item $\sigma^{-2}(\Delta\gamma)$:  This number indicates the ability of a
program to test that the growth of structure follows that expected in GR.
Larger values indicate greater power.  Our data models for pre-JDEM results
indicate a value of $\sigma(\Delta\gamma)=0.21$. A JDEM should result in a
smaller number.  We note that $\sigma(\Delta\gamma)$ is primarily a test of GR
at $z<1$, as the FoMSWG cosmological model requires deviations from GR growth
to be much larger at low $z$.

\end{enumerate}

\appendix \label{appa}

\section{Predictions for the pre-JDEM Fisher matrices}

\subsection{Planck Fisher Matrix}

CMB observations are valuable dark-energy studies because they (i) enable
us to estimate the matter density $\omega_M$ and the distance
$D_{\rm A}(z=1100)$ to the surface of last scattering; (ii) set the
amplitude and shape of the matter power spectrum and length of the BAO
standard ruler; and (iii) contain secondary anisotropies that carry
low-redshift information such as the ISW effect, SZ effect, and lensing of
the CMB.  Accordingly, the FoMSWG constructed a Fisher matrix intended to
model dark energy constraints from Planck.  This Fisher matrix only
contains the primary anisotropies; dark energy constraints from, e.g., the
SZ effect would be considered part of the cluster technique.  Its method
of construction is similar to that of the DETF Planck matrix, but is not
exactly the same.

The Fisher matrix was constructed using the Planck LFI 70 GHz and HFI 100
and 143 GHz channels using the ``performance goal'' beam and noise
parameters from the Planck Blue Book \cite{2006astro.ph..4069T}.  Both
temperature and $E$-mode polarization data were used over 70\% of the sky.
It was assumed that over this fraction of the sky the other channels could
be used to remove foregrounds, and that the Planck beams could be mapped
well enough that they would not be a limiting systematic.

In most cases, the Fisher matrix is constructed using the standard
equation,
\begin{equation}
\mathcal{F}_{ij} = \frac12\sum_{\ell=2}^{2000} (2\ell+1)f_{\rm sky}{\rm Tr}
\left[
({\bf C}_\ell+{\bf N}_\ell)^{-1}\frac{\partial{\bf C}_\ell}{\partial p^i}
({\bf C}_\ell+{\bf N}_\ell)^{-1}\frac{\partial{\bf C}_\ell}{\partial p^j}
\right],
\end{equation}
where $i$ and $j$ represent the cosmological parameters, $\ell$ is multipole
number, $f_{\rm sky}=0.7$ is the fraction of the sky observed, ${\bf C}_\ell$
is the $2\times 2$ CMB power spectrum matrix (containing $TT$, $TE$, and
$EE$ spectra), and ${\bf N}_\ell$ is the $2\times2$ noise matrix.

The parameter space is the standard FoMSWG parameter space, augmented by the
optical depth $\tau$ due to reionization.
Two exceptions are made to the above rule.
\begin{enumerate}
\item
For the cosmological parameters $X\in\{\Omega_{\rm K},w_0...w_{35}\}$, we
force the exact angular diameter distance by making the replacement:
\begin{equation}
\frac{\partial{\bf C}}{\partial X} \leftarrow \frac{\partial D_{\rm
A}(z=1100)/\partial X}
{\partial D_{\rm A}(z=1100)/\partial\omega_{\rm de}}
\frac{\partial{\bf C}}{\partial\omega_{\rm de}}.
\end{equation}
This exactly enforces the angular diameter distance degeneracy and removes
any information from the power spectrum of the ISW effect.
\item
The optical depth $\tau$ is unique in depending mainly on the lowest
multipoles, even with Planck instead of WMAP.  Observations have shown
that these low CMB multipoles are below Galactic foregrounds in
polarization over most of the sky at all frequencies
\cite{2007ApJS..170..335P}, and reaching the Planck statistical   
uncertainty of $\sigma(\tau)\sim 0.005$ may be too optimistic.  Moreover,
the single-parameter step models of reionization are undoubtedly too
simple and the ability to measure $\tau$ is degraded when one considers   
more generic possibilities.  It is estimated that these uncertainties may
limit Planck constraints on optical depth to about $0.01$ 
\cite{2003ApJ...595...13H}.  We thus eliminated the $\ell<30$ multipoles in
the $TE$ and $EE$ spectra (i.e., we set $N^{EE}_\ell=\infty$) and replaced 
them with a prior on $\tau$ with $\sigma(\tau)=0.01$.  This prior dominates
over any information about $\tau$ from the $TT$ power spectrum.
\end{enumerate}

We did not include primordial gravitational waves (``tensors'') in the Fisher
matrix.  A Fisher matrix calculation shows that over $f_{\rm sky}=0.7$, a
tensor-to-scalar ratio of $r=0.03$ is required to produce a $1\sigma$ change in
the CMB $TT$ power spectrum and $r=0.06$ to produce a $2\sigma$
change.\footnote{Our normalization  convention corresponds to $r=16\epsilon$ in
terms of the inflationary slow-roll parameter.} On the timescale of JDEM, it is
likely that $B$-mode observations will have reached or approached this level of
precision. Moreover, tensors mainly affect the low multipoles ($\ell<100$)
whereas most of the Planck constraints relevant to dark energy come from higher
multipoles (the exception, noted above, is $\tau$).  Thus at the level of
accuracy of this report, it is reasonable to neglect tensors in dark energy
forecasts even though they are well-motivated.

Since $\tau$ does not appear in any of the non-CMB cosmological probes, we
marginalize it out of the Planck Fisher matrix before combining it with   
other information.

\subsection{Pre-JDEM Weak Lensing Fisher Matrix}

The FoMSWG projection of weak lensing data in ca.\ $2015$ is based on
estimates of the performance of the {\em Dark Energy Survey (DES)},
which has been recommended for CD-3b approval from the Department of Energy
and plans to commence its survey around 2011 and to complete around 2016.
A shallower, larger-area WL survey may be completed by this date by the
{\it PanSTARRS} collaboration, but the available information was
insufficient to evaluate its performance.

Our code is based on the DETF model, but with some improvements, most notably
more thorough treatment of intrinsic alignments (described in detail in
\S\ref{subsec:ch}).

\subsubsection{Survey parameters}

For the DES we assume a sky coverage of 5000~deg$^2$, and a total effective
source-galaxy density $n_{\rm eff}=9\,{\rm arcmin}^{-2}$ with shape noise per
galaxy of $\sigma_\gamma=0.24$.  We presume the sources to be distributed in
redshift according to $dn_{\rm eff}/dz \propto z^2 e^{-z/z_0},$ with $z_0$
chosen to yield a median source redshift of 0.6.  Photometric redshifts errors
for DES are presumed to have a Gaussian distribution with
$\sigma_z=0.05(1+z)$.  We ignore redshift outliers.

We divide the source population into 10 bins, sufficient to saturate the
dark-energy information content.  We consider information to be available from
the 2-point correlations between all bin pairs.  Each bin contains both
galaxy-density and galaxy-shear observations.  We include information from
shear-shear correlations (i.e., cosmic shear power spectrum) and shear-density
correlations (the cross-correlation cosmography test). We have neglected
non-Gaussian observables (3- and 4-point functions, WL peak or cluster counts,
aperture mass PDFs, etc.) due to the difficulty of forecasting their systematic
errors.

These codes consider all 2-point observables, plus systematic errors
due to:
\begin{list}{$\bullet$}{}
\item shear calibration errors;
\item photo-$z$ biases;
\item intrinsic alignments of galaxies (IA), both shape-shape
  correlations (II) and shape-density correlations (GI);
\item weak prior constraints on the biases and correlation
  coefficients of galaxy density and intrinsic alignments with the 
  mass distribution; and
\item non-Gaussian covariance of convergence power spectra.
\end{list}

\subsubsection{Formalism}
\label{subsec:ch}

Our code has not been described previously; there is substantial overlap with
DETF but for purposes of reproducibility it is described here in detail.

We divide the source galaxies into $N_z=10$ redshift slices ($z$-bins) and
consider power spectra in each of $N_\ell=18$ multipole bins ($\ell$-bins).  
The $z$-bins have equal numbers of galaxies; their redshift centroids range from
0.18 (1st bin) to 1.41 (10th bin).  The $\ell$-bins are logarithmically spaced
between $\ell=12$ (1st bin) and $\ell=7700$ (12th bin). In each angular bin we
construct all power spectra and cross-spectra of the galaxy densities
$\{g_i\}_{i=1}^{N_z}$ and convergences inferred from shear maps
$\{\kappa_i\}_{i=1}^{N_z}$.  There are $N_z(2N_z+1)$ such power spectra and
cross spectra, and so we construct the length $N_\ell N_z(2N_z+1)$ vector of
power spectra ${\bf C} = \{C^{\kappa\kappa}_{\ell ij}, C^{gg}_{\ell ij},
C^{g\kappa}_{\ell ij}\}$.  A Fisher matrix requires (i) a model for ${\bf C}$ as
a function of cosmological and nuisance parameters, and (ii) a $N_\ell
N_z(2N_z+1) \times N_\ell N_z(2N_z+1)$ covariance matrix $\Sigma$.  A systematic
error may be included as either a nuisance parameter (possibly with a prior), or
as an additional contribution to $\Sigma$. Here shear calibration and photo-$z$
biases are included as nuisance parameters, and other errors (galaxy biasing and
intrinsic alignments) are included in $\Sigma$.

\subsubsection{Power spectrum model}

In order to estimate the angular galaxy and convergence power spectra, we need
the 3-dimensional power spectrum matrix of the matter $m$, galaxies $g$, and
convergence reconstructed from intrinsic ellipticity $e$:\footnote{In general
$e$ depends on galaxy shapes and hence on the line of sight along which the
galaxies are observed; since we will use the Limber approximation one should
assume that this line of sight is perpendicular to ${\bf k}$.}
\begin{equation}
{\bf P}(k,z) = \left( \begin{array}{ccc} P_{mm} & P_{gm} & P_{em} \\ 
P_{gm} & P_{gg} & P_{ge} \\ P_{em} & P_{ge} & P_{ee} \end{array} \right).
\end{equation}
(For clarity, we suppress $k,z$ indices where not needed.)  This matrix is
symmetric because rotational invariance allows one to transform ${\bf
k}\rightarrow -{\bf k}$, and in general contains 6 power spectra.  We include
shot noise in $P_{gg}$ but not $P_{ee}$.\footnote{This is so that we may use the
stochasticity $r_g$ as an estimate of how well the galaxy density can be used to
trace the mass, but also set $P_{ee}=0$ in our fiducial model.}

All 6 of these spectra need to be modeled, but only the matter power spectrum
$P_{mm}(k,z)$ is known accurately from theory. We use the Eisenstein and Hu
transfer function \cite{EisensteinHu} and the Smith et al.\ nonlinear mapping
\cite{Smithetal}.  Unlike DETF, we take into account the full $\Omega_{\rm m}$
dependence of the nonlinear mapping for consistency, although its validity for
arbitrary $w(z)$ still needs to be studied.

The galaxy spectra are controlled by a bias $b_g$ and stochasticity $r_g$:
$P_{gg} = b_g^2 P_{mm}$ and $P_{gm} = b_gr_gP_{mm}$.  Both of these can in
general depend on scale and redshift. 
Similarly for the intrinsic ellipticity, one may define $P_{ee}=b_\kappa^2
P_{mm}$ and $P_{me} = b_\kappa r_\kappa P_{mm}$.  This leaves the $P_{ge}$
spectrum undetermined; no specific model is required since we
marginalize over it. The fiducial model contains no intrinsic alignments, i.e.,
$P_{ee}=P_{ge}=P_{me}=0$.

The convergence power spectra are given by the standard Limber equation:
\begin{eqnarray}
C_{\ell ij}^{\kappa\kappa} &=& \frac9{16}(\Omega_{\rm m}H_0^2)^2(1+f_i)(1+f_j)
\int (1+z)^2 A_i(z) A_j(z) P_{mm}(k,z)\frac{dz}{H(z)}
\nonumber \\
&& + \delta_{ij} P_{ee}(k_{(i)},z_i) \frac{H(z_i)}{\Delta z_i}
\nonumber \\
&& +\frac34\Omega_{\rm m}H_0^2(1+f_j)
\frac{A_j(z_i)}{[r(z_i)]^2}
P_{em}(k_{(i)},z_i) + (i\leftrightarrow j)
\nonumber \\
&& + \frac{\gamma_{\rm rms}^2}{n_i}\delta_{ij}.
\label{eq:kk}
\end{eqnarray}
In this equation, the first line represents the pure lensing part of the power
spectrum, the second line represents II contamination, the third GI
contamination, and the fourth shot noise.  We define $k=\ell/r(z)$ where $r(z)$
is the comoving angular diameter distance, $n_i$ the source density in the $i$th
bin, and the lensing strengths are defined by
\begin{equation}
A_i(z) = 2\frac{r(z_i,z)r(z)}{r(z_i)}\Theta(z_i-z),
\end{equation}
where $r(z_i,z)$ is the comoving angular diameter distance from the lens
redshift $z$ to the source redshift $z_i$ (i.e., the redshift of the galaxies in
the $i$th redshift bin).  $\Theta$ represents the Heaviside step function.

The parameters $f_i$ in Eq.~(\ref{eq:kk}) are the shear calibration parameters. 
They are zero for perfect shear estimation; in reality one will have to apply a
prior with mean zero and some uncertainty $\sigma(f_i)$.  Note that $\chi$, $k$,
and $A$ contain implicit cosmology dependence.

Similarly, the galaxy power spectra are given by
\begin{equation}
C_{\ell ij}^{gg} = \delta_{ij} P_{gg}(k_{(i)},z_i) \frac{H(z_i)}{\Delta z_i},
\label{eq:gg}
\end{equation}
where $k_{(i)}=\ell/r(z_i)$ and $\Delta z_i$ is the width of the $i$th bin.  The
galaxy power spectra are not actually used to probe cosmological parameters,
however the power spectra are needed in order to construct $\Sigma$.
The galaxy-convergence spectra are
\begin{equation}
C_{\ell ij}^{g\kappa} = \frac34\Omega_{\rm m}H_0^2(1+f_j)
\frac{A_j(z_i)}{[r(z_i)]^2}
P_{gm}(k_{(i)},z_i)
+ \delta_{ij} P_{ge}(k_{(i)},z_i) \frac{H(z_i)}{\Delta z_i},
\label{eq:gk}
\end{equation}
which is zero for $j<i$ and measures intrinsic alignments for $j=i$.

The shell redshifts $z_i$ are assigned fiducial values so that equal numbers of
source galaxies are in each bin, and priors $\sigma(z_i)$ are assigned on the
knowledge of the photo-$z$ bias.

\subsubsection{Covariance matrix model}

In the case of Gaussian density and convergence fields, the covariance matrix
takes the form of Wick's theorem:
\begin{equation}
\Sigma[C^{AB}_{\ell ij}, C^{A'B'}_{\ell' i'j'}] = 
\frac1{2f_{\rm sky}\ell^2\Delta\ln\ell}
  \left(C^{AA'}_{ii'}C^{BB'}_{jj'} + C^{AB'}_{ij'}C^{BA'}_{ji'}
  \right)\delta_{\ell\ell'},
\end{equation}
where $A$ and $B$ represent either $g$ or $\kappa$, and the denominator in the
prefactor counts the number of modes in a bin.

We include observables for $\ell<10^4$, but include the non-Gaussian covariances
due to the 1-halo term\footnote{We do not yet include the dispersion in the
concentration-mass relation, which may be significant at the 4-point level
\cite{2001ApJ...554...56C}.  Neither do we include halo triaxiality (which
boosts the dispersion in projected concentrations).} in the matter 4-point
function, which has been shown to dominate the excess covariance of the power
spectrum on small scales \cite{2001ApJ...554...56C}.  The galaxy ($gggg$) and
galaxy-matter ($ggmm$) trispectra are more complicated to model and not yet
included; while we expect that trispectrum effects will be less important for
the cross-correlation cosmography ratio tests where one does not use the
amplitude of the galaxy-matter correlation, this issue deserves future study. 
The specific equation for this 4-point contribution is
\begin{eqnarray}
\Sigma[C^{\kappa\kappa}_{\ell ij},C^{\kappa\kappa}_{\ell'i'j'}] 
&+=& \frac{81}{256\pi^2f_{\rm sky}}(\Omega_{\rm m}H_0^2)^4 \int dz\,(1+z)^4
\int dM\, \frac{M^4}{[r(z)]^8}
\nonumber \\ && \times
\frac{dN}{dz\,dM} [u(k,z,M)u(k',z,M)]^2
\nonumber \\ && \times
A_i(z) A_j(z) A_{i'}(z) A_{j'}(z),
\end{eqnarray}
where ``$+=$'' means the adding of the right-hand side to the specified
covariance matrix element.  Here $dN/dz\,dM$ is the number of halos in the
whole sky per unit redshift $z$ per unit mass $M$, and $u$ is the Fourier
transform of the halo profile at comoving wavenumber $k=\ell/r(z)$ or
$k'=\ell'/r(z)$, normalized to $u(k=0)=1$.  We use the Sheth \& Tormen
\cite{1999MNRAS.308..119S} mass function and an NFW \cite{1997ApJ...490..493N}
mass profile.  The concentration-mass relation was taken from the appendix of
Voit et al.\ \cite{2003ApJ...593..272V}, rescaled to the Sheth \& Tormen
definition of virial overdensity.

We next marginalize out the galaxy biasing and intrinsic alignment
parameters.  There are five of these.  The galaxy power spectra $P_{gg}$ are
easy to marginalize out: one removes all information about $C^{gg}_{\ell ij}$:
\begin{equation}
\Sigma[C^{gg}_{\ell ij}, C^{gg}_{\ell ij}] += \mu,
\end{equation}
where $\mu\rightarrow\infty$
for all $\ell$, $i$, $j$.\footnote{Technically the $i\neq j$ galaxy 
cross-powers need not be marginalized out because they should be zero in the 
fiducial model; however we assume that they will be used to constrain photo-$z$
outlier models and hence should not be used again in the determination of 
cosmological parameters.}

Any remaining nuisance parameter $X$ associated with the galaxy biasing or
intrinsic alignments is taken out by setting:
\begin{equation}
\Sigma[C^{AB}_{\ell ij}, C^{A'B'}_{\ell' i'j'}] += 
\sigma^2_{\rm prior}(X)\frac{\partial C^{AB}_{\ell ij}}{\partial X} 
\frac{\partial C^{A'B'}_{\ell'i'j'}}{\partial X},
\end{equation}
where $\sigma^2_{\rm prior}(X)$ is the prior placed on parameter $X$.  We
marginalize over $P_{gm}(k,z)$ in each of our $\ell$ and $z$-bins (i.e., $N_\ell
N_z$ parameters total) with sufficiently large priors that the priors have no
effect on the result.  (They cannot be infinite for numerical reasons.)  This 
nonparametric marginalization prevents our Fisher matrix from taking advantage
of any functional form we specify.  A similar method is used for $P_{ee}(k,z)$
(which effectively removes shear auto-powers within the same bin) and
$P_{ge}(k,z)$.  For $P_{me}(k,z)$, which has the most significant degrading
effect on cosmological parameters, we implement a prior in each bin to
prevent unrealistically large intrinsic alignments.  This prior is described and
justified in \S\ref{subsec:ia}.

\subsubsection{Shear calibration errors}

Shear calibration errors are multiplicative errors on the shear $\gamma$:
$\gamma \rightarrow (1+f)\gamma$.  The calibration error $f$ must be taken to
depend on redshift $z$ because as a function of redshift one observes galaxies
of different types (morphology, radial profile) and colors (hence different
PSF!).  The STEP simulations \cite{2007MNRAS.376...13M} have shown that in
simulations with a wavelength ($\lambda$)-independent PSF, several shape
measurement methods currently achieve shear calibration better than 2\%
($|f|<0.02$).  With further STEP-like exercises with simulated galaxies that
closely resemble the observed morphologies of the actual DES sample, we
anticipate further reductions in $|f|$.  In the near future, the shear
calibration may become limited by $\lambda$-dependent PSF effects: galaxies
whose light is dominated by longer wavelengths with smaller PSFs will have
larger calibration than galaxies dominated by short wavelengths.  The lensing
community is in the early stages of thinking about this problem, so we have
not assumed that advanced solutions\footnote{This would include, for example,
decomposing multicolor galaxy images into star-forming and old stellar
population components, convolving with model SEDs for each component and
fitting a shape.} will be available for pre-JDEM.

We noted that for a ground-based system with seeing limited by
Kolmogorov-spectrum turbulence in the atmosphere, the PSF 2nd moments scale as
$\lambda^{-2/5}$.  For a galaxy with typical resolution factor of about $0.5$
(as defined in Hirata \& Seljak \cite{2003MNRAS.343..459H}) the shear
calibration changes by 0.4\% for every 1\% change in effective wavelength.  The
RMS fluctuation in wavelength for the extreme case of an emission line sitting
in a filter of width $\Delta\lambda/\lambda = 0.25$ is $0.25/\sqrt{12}=0.072$,
corresponding to a shear calibration fluctuation of 0.029.  A single emission
line causes the worst damage, but most galaxies in most filters have only a
small fraction of their emission in lines. As a first estimate of the more
realistic case, we took 20\% of the single-line case, assigning a shear
calibration error of $0.0058$.\footnote{The effective wavelength of the PSF can
also be affected by breaks in the spectrum.  For a galaxy with a factor of 2
break in the SED in the center of the bandpass, the effective wavelength is
shifted by 2\% from a flat SED, corresponding to a shear calibration
fluctuation of 0.008.} We take one bin ($\Delta \ln a \sim 0.1$) as a
reasonable correlation length for the shear calibration systematic because a
galaxy redshifts through a filter in $\Delta \ln a \sim 0.25$.  A systematic
that depends on e.g., an emission line appearing in the red or blue side of a
filter should change sign twice during a filter passage.  Therefore a prior of
$\sigma(f_i)=0.0058$ was applied independently in each $z$-bin.

\subsubsection{Photometric redshifts}

We assume that in about $2015$ there will be available an {\em unbiased}
spectroscopic redshift survey of order $10^5$ galaxies down to the DES limit of
$I_{\rm AB}\approx23.5$~mag. With this size spectroscopic survey, the ability to
constrain the photo-$z$ error properties will be dominated by two systematic
errors: (i) incompleteness due to redshift failures, and (ii) photometric
calibration or extinction fluctuations in the lensing survey (such that the
spectroscopically targeted regions are not representative of the whole).  These
cannot be fully assessed with the information available and without knowing the
extent of the resources that will be brought to bear on the photo-$z$ problem. 
A simple calculation shows that losing 5\% of the galaxies, all on one tail of
the distribution, would lead to a bias of $0.005(1+z)$ in the resulting
photo-$z$ error distribution.  A bias of the same order of magnitude might be
expected to arise from about a $1$\% photometric calibration fluctuations
(suppressed by a factor of a few due to the spectroscopic survey covering
several patches in the larger WL survey area).  

The ``nightmare'' scenario is the existence of a class of photo-$z$ outliers
with no features strong enough to identify a spectroscopic redshift.  While
cross-correlation techniques \cite{2008ApJ...684...88N} should be able to
identify this problem, one is left with the ambiguity in the bias of the
outliers since angular correlations only measure the product of bias and
redshift distribution $b\,dN/dz$ \cite{2008PhRvD..78d3519H} (with corrections
due to stochasticity on nonlinear scales).  We have assumed that the combination
of spectroscopic surveys and galaxy correlations can constrain the photo-$z$
bias to a $1\sigma$ error of $0.005(1+z)$, but this number is in urgent need of
refinement.

\subsubsection{Galaxy biasing and cross-correlation cosmography}

In WL surveys, there is potentially much to gain in using the galaxy density as
an indicator of the underlying mass distribution.  An example is the use of the
source redshift dependence of galaxy-shear correlations in lensing
cross-correlation cosmography \cite{2003PhRvL..91n1302J, 2004ApJ...600...17B}. 
This is true even if the bias $b_g$ and stochasticity $r_g$ of the galaxy
density are poorly known. The gain from this approach depends on the correlation
coefficient of galaxy density $r_g$ with respect to mass.  We assume a fiducial
model in which $r_g=0.9$ in the linear regime, dropping to $r_g=0.6$ in the
non-linear regime.

Galaxy density information is completely rejected by
marginalizing over $b_g$ and $r_g$ in each of the 10 redshift bins and
18 angular-scale bins.  Note that this procedure preserves information
in the lensing distance ratio test.

\subsubsection{Intrinsic alignments}
\label{subsec:ia}

We take a fiducial model with no intrinsic alignments, so that it is impossible
for intrinsic alignments to be used to predict  the local mass distribution. 
If $b_\kappa$ is large and the IA model  is sufficiently restrictive, it is
possible for the Fisher matrix to  take advantage of the cosmology dependence
of the intrinsic alignments,  which is an undesired result
\cite{2007NJPh....9..444B}.

Intrinsic alignment models must be allowed to be functions of redshift and of
angular scale in a realistic forecast. GI and II effects are treated
separately.  GI depends on the matter-IA  correlation ($b_\kappa r_\kappa$)
whereas II depends on $b_\kappa^2$.   We completely marginalize over II, i.e.,
we allowed $b_\kappa^2$ to float  independently in each ($z,\ell$) bin.  This 
is equivalent to rejecting  the shear power spectra within a $z$-slice and only
taking cross-spectra  \cite{2004ApJ...601L...1T}; this may be conservative but
little  information is lost through complete marginalization over II.  In the 
case of GI, we assumed that in the linear regime (crudely modeled as 
$\ell<300$) galaxies trace the matter well enough that the 
galaxy-intrinsic ellipticity correlation could be used to infer the 
matter-intrinsic alignment correlation $b_\kappa r_\kappa$ and apply the 
corresponding small correction to the lensing data (expected to be around
$3\sigma$ according to the Hirata et al.\ IA model B 
\cite{2007MNRAS.381.1197H}).  On nonlinear scales, we applied a weak  prior on
$b_\kappa r_\kappa$ to prevent it from being much larger than  the observed IA
signal of $|b_\kappa r_\kappa|\sim 0.003$  \cite{2006MNRAS.371..750H}.  In the
absence of knowledge of the  ``correct'' form for such a prior, we used
\begin{equation}
\sigma(b_\kappa r_\kappa) = 0.003\sqrt{N_{\ell,\rm nonlin}(N_z-1)},
\end{equation}
applied independently in each $(z,\ell)$ bin in the nonlinear regime.  
This choice was motivated by requiring that we not be able to constrain 
$b_\kappa r_\kappa$ to better than 0.003 by 
``averaging down'' the prior in the different $(z,\ell)$ bins.  We use 
$N_z-1$ instead of $N_z$ in the above formula because $b_\kappa 
r_\kappa$ in the highest-$z$ bin has no effect on the observed power 
spectra.

The 0.003 number is highly uncertain as it results from an extrapolation 
of SDSS data to moderate redshift.  It will have to be revisited after 
either SDSS spectroscopic intrinsic alignment analyses 
\cite{2007MNRAS.381.1197H} can be repeated on higher-redshift samples, 
or photometric WL surveys such as DES provide a refined estimate of the 
order of magnitude of $b_\kappa r_\kappa$.  A particularly nasty (but 
possible) scenario would be that galaxy alignments are strong during the peak
era of galaxy formation at $z>1$, but decay at later times as galactic 
disks precess and de-align in their dark matter halos.  In this case 
the intrinsic alignment signal even at $z\sim 0.6$ could be much 
stronger than in SDSS and we may be forced to the pessimistic option of 
completely marginalizing over $b_\kappa r_\kappa$ in the nonlinear 
regime.

\subsection{Pre-JDEM Baryon Acoustic Oscillation Fisher Matrix}

Forecasts for the baryon acoustic oscillations were computed using
the formulae from Seo and Eisenstein \cite{Seoa}.  This produces fractional
errors on $D_A(z)/s$ and $H(z)s$, where $s$ is the sound horizon and
varies with $\Omega_m h^2$ and $\Omega_b h^2$.  In all cases, the 
errors on $D_A/s$ and $Hs$ are mildly covariant, with a coefficient
of 0.4 (in the sense that a larger $D_A$ tends to have a larger $H$).

We used redshift bins identical to the $w(z)$ parametrization.
Each bin was said to yield an independent constraint on the distance
scale.  The comoving volume in each bin was used to scale the answer
from the large volume limit; i.e., no boundary effects were considered.
This should be a reasonable approximation for large surveys and
slow variations in $w(z)$.  Note that although $w(z)$ was given a 
very flexible parametrization, the cosmological constraints are 
such that only slow variations are well measured.

For simplicity, each survey was taken to have a constant comoving
number density $n$ and a constant amplitude of the power spectrum
$P_{0.2}$.  This is not exact but it is hard to see how it could 
alter the results at any level that would affect a JDEM choice.
For convenience, we registered the redshift range of the surveys
to the redshift binning.  The power spectrum was taken to be
anisotropic according to the linear Kaiser model, with $\beta$
computed from $f=\Omega_m^{0.6}(z)$ and the bias.

It should be noted that these are not exact models of the surveys,
but they do capture the scope and redshift range of the projects.

No systematic errors were included.  Simulations show no evidence
for systematic errors in the acoustic scale at the level of precision
of these surveys, nor are survey selection effects expected to pose
any insurmountable problems \cite{Seob,Sanchez}.  

The degradation of the acoustic scale due to non-linear structure
formation was included assuming that $\Sigma_\perp = 9(D/D_0)h^{-1}$~Mpc,
where $D$ is the growth function and $D_0$ is that function at
$z=0$.  The transverse degradation was $\Sigma_\parallel = \Sigma_\perp(1+f)$.

Density-field reconstruction was included as a multiplicative
reduction in these two non-linear parameters.  The best-case 
multipliers were taken to be 0.5, but some sparser samples were
given 75\% or even 100\% of the non-linearity.

The surveys employed are listed in Table \ref{BAOPROJECTS}.

\begin{table*}[!ht]
\begin{ruledtabular}

\begin{tabular}{llllcc}

Survey & Redshift\footnote{Redshift ranges picked to match the bin boundaries
in $w(z)$, hence the non-standard values.} & Area & $\sigma_8$ & \# of galaxies
& Non-linearity \\ \hline \hline

WiggleZ & $0.6<z<1.0$ & 1000 deg$^2$ & 0.8 & 0.4M & 75\% \\

BOSS & $0.08<z<0.48$ & 10,000 deg$^2$ & 1.5 & 0.8M & 50\% \\

BOSS & $0.48<z<0.67$ & 10,000 deg$^2$ & 1.5 & 0.4M & 50\% \\

BOSS Ly-$\alpha$  & $2.08<z<2.64$ & 8000 deg$^2$ & 0.25 & 
7.5M\footnote{Number of ``galaxies" picked to give aggregate $D$ error of 1.7\% 
and $H$ error of 1.5\%, so as to match the BOSS Lyman $\alpha$ forest forecast.}
& 100\%  \\

WFMOS & $0.60<z<1.0$ & 2000 deg$^2$ & 0.8 & 2M & 50\% \\

WFMOS & $1.00<z<1.35$ & 2000 deg$^2$ & 0.8 & 1.5M & 50\% \\

HETDEX & $1.86<z<3.44$ & 420 deg$^2$ & 0.8 & 0.8M & 50\% \\

Photo-$z$ surveys\footnote{4\% fractional error in $1+z$} & $0.67<z<1.10$ 
& 20,000 deg$^2$ & 1.4 & 10M & 100\% 

\end{tabular}
\end{ruledtabular}
\caption{A description of the BAO surveys assumed in the FoMSWG pre-JDEM data
model.\label{BAOPROJECTS}}

\end{table*}

We omitted the high-redshift WFMOS data set arbitrarily as we were including
BOSS and HETDEX at high redshift.  Including any two of the three planned
experiments gives similar results.

These distance estimates include only the results from the BAO scale.  It is
likely that these redshift surveys will yield other cosmological information
relevant to dark energy, e.g., the redshift distortion and Alcock-Paczynski
methods described in Sec.\ \ref{sec:AP}. Both require detailed modeling of
non-linear structure formation and galaxy clustering bias.  If this modeling
can be done accurately, the dark energy information could be substantially
improved, and this promise is driving considerable research on these topics.

\subsection{Pre-JDEM Supernova Fisher Matrix}

\subsubsection{The Importance of Systematic Uncertainties}

The best-established method for measuring the history of cosmic expansion is
the use of luminosity distances to supernovae (Type Ia).  Our goal here is to
estimate the sample of supernova observations that will be available at the
time of a JDEM launch, taken to be 2016, so that we can measure the incremental
impact of a JDEM on knowledge of cosmological parameters using our
figure-of-merit. Our conclusion detailed below, is that making reasonable
assumptions about observing programs that have, and have not yet been carried
out, the samples at all redshifts below 0.8 will be large enough to make
statistical errors irrelevant, provided there is no stringent reduction of the
usable number of objects by selecting for a narrow set of host galaxy or
supernova properties. This means that systematic errors, that are not reduced
by increasing the sample size, will dominate.  Such systematic errors are
difficult to enumerate and measure.  The only certain way to see their effect
is to compare independent results from completely different techniques.  The
task ahead for supernova cosmology is to explore the best ways to measure and,
if possible, reduce these distance errors.  The task for a JDEM proposer will
be to describe their program for improving on what will have been done by 2016
in the statistical language of the FoMSWG.  We caution that making less
realistic assumptions about systematic errors is the simplest way to show
"improvement" in the figure of merit.  If used in this way, the figure of merit
could reward the least realistic and punish the most thoughtful proposal.

What are the systematic errors that keep today's distance measurements from
achieving the precision set by statistical error?  There are many.  These
include photometric calibration across a factor of 10 in wavelength from the
U-band to the near infrared, and a factor of $10^6$ in flux from the nearest
supernovae (in the Hubble flow) to the most distant objects detected by HST.
They include peculiar velocities and regional flows that may affect the correct
interpretation of the velocity-distance relation.  They include uncorrected
selection effects and genuine evolution in the mean properties of the nearby
and distant samples.

But most importantly, they include errors that result from the way in which the
light curves are used to infer the distances.  The rise and decline of SN Ia
light curves are closely linked to the intrinsic luminosity of the supernova:
measuring the shape of the light curve allows the uncertainty in distance for a
single supernova to be reduced from roughly 50\% to under 10\%.  An additional
source of uncertainty comes from the effects of interstellar dust, both in our
Galaxy and in the supernova host.   Interstellar dust does not absorb equally
at all wavelengths. The current evidence from supernovae \cite{Conley.2007}
shows that the ratio of total absorption to absorption at a particular color
that gives the smallest dispersion in supernova results is not the value
usually associated with dust in our Galaxy.  The net result is that inferences
about cosmological parameters (such as $w$) are systematically different
depending on which light curve fitter is used, and which method of accounting
for the effect of dust is employed.  Estimates of the magnitude of these
effects vary, and are generally in the range of 7-10\%, but there is currently
no published result where the systematic effect on $w$ is claimed to be smaller
than 6\% . External validation of systematic uncertainty estimates would demand
independent sets of low- and high-redshift data and a comparison of the results
obtained with a variety of approaches to light curve fitting and extinction
corrections.

One path to avoiding the systematics of dust and evolution is to discard a
large fraction of the discoveries, while retaining only those in elliptical
galaxies.  This requires a careful demonstration that the supernovae in
ellipticals can give a better cosmological result and confidence that the dust
absorption in ellipticals is fully understood.

A possible path to better understanding of the dust that obscures and reddens
supernovae may come from observations made in the rest-frame near infrared
(J,H,K bands).  Preliminary indications are that SN Ia are more similar to one
another at infrared wavelengths than in the optical bands, and the effects of
dust are smaller by about a factor of 3 \cite{Wood-Vasey.2007b}.  The combined
analysis of optical and infrared data has the potential to determine the dust
properties for individual objects \cite{Krisciunas.2007} and to place this
subject on firmer footing.

In summary, we are reluctant to assume that these unsolved systematic problems
in supernova cosmology will evaporate in the next 7 years (before JDEM).
Therefore, we assume a conservative floor to the distance uncertainties,
amounting to 2\% in distance in each bin of width $\Delta z=0.1$ for $z\le0.5$,
and 3\% in distance  in each bin of width $\Delta z=0.1$ for $z\ge0.5$.  The
systematic distance errors are taken to be statistically independent from bin
to bin.  The discussion below will show that present and future SN surveys are
likely to detect enough SNe to reduce statistical distance errors below these
systematic floors for all $z<0.9$.

In the computation of the FoMSWG figure of merit example for the most
significant principal components, the assumed JDEM sample was that used by the
DETF.

\subsubsection{Assumptions about pre-JDEM SNe Surveys}

The early results of the High-Z Team \cite{Riess.1998} and the Supernova
Cosmology Project \cite{Perlmutter.1999} provided clear evidence for cosmic
acceleration, and, because they used different approaches for using the light
curve shape to determine the absolute magnitude, as well as different samples of
high redshift supernova, provided some external tests of the estimated errors.
The DETF made some forecasts of the available data for Stage II and Stage III
supernova samples.  Since then, the samples at high redshift ($z \sim 0.5$) have
been substantially augmented, especially by the ESSENCE program
\cite{Wood-Vasey.2007a}, the Supernova Legacy Survey \cite{Astier.2006} and at $
z > 1$, in a development not considered by the DETF, by the Higher-Z program on
HST \cite{Riess.2004, Riess.2007}.  The published samples at low redshift, which
are equally important in deriving the cosmology, have also been augmented by SN
Ia light curves from the CfA \cite{Jha.2007} and the SCP \cite{Kowalski.2008}.
Another important development, not anticipated by the DETF, has been the SDSS
Supernova Survey, which has good light curves (better than 5 epochs, with one
before maximum) and spectra for confirming the supernova type and determining
its redshift for 300 objects in the redshift range $ 0.1 < z < 0.3$.

\begin{table*}[!ht]
\begin{tabular}{lr}
\hline\hline
redshift bin & number of SNIa 	\\
\hline \hline
$<0.1$	&	500		\\			
0.1-0.2	&	200		\\			
0.2-0.3	&	320		\\
0.3-0.4	&	445		\\
0.4-0.5	&	580		\\
0.5-0.6	&	660		\\
0.6-0.7	&	700		\\
0.7-0.8	&	670		\\
0.8-0.9	&	110		\\
0.9-1.0	&	80		\\
1.0-1.1 &     	25		\\
1.1-1.2 &      	16		\\
1.2-1.3	& 	16		\\
1.3-1.4	&   	4		\\
1.4-1.5	&   	4		\\
1.5-1.6	&   	4		\\
$>1.6$	&   	4		\\
\hline\hline
\end{tabular}
\caption{The number of SNIa in various redshift bins assumed in constructing 
the pre-JDEM data model.\label{SNBINS}}
\end{table*}

Looking ahead, we note that the low-redshift sample is on the verge of rapid
expansion.  In the past year, the SCP added 8 objects to the low-$z$ data set
\cite{Kowalski.2008}.  The CfA SN program has 150 more, most of which will
pass the same data cuts, the Carnegie SN program promises 200 light curves, and
the KAIT program at the Lick Observatory has approximately 150 to contribute.
While there may be some overlap among these objects, the DETF assumption of 500
well-observed nearby objects seems reasonable.  Skymapper, a survey telescope
in Australia, will create a well-sampled and carefully controlled set of 150
low-$z$ supernovae, once it begins operation in 2009. At redshifts above 0.1,
there will be no extension of the SDSS Supernova Survey and no continuation of
the Supernova Legacy Survey.  However, there will be the advent of a supernova
survey that follows up the ``medium-deep'' fields of Pan-STARRS.  This should
begin in 2009.  We assume it will run for 40 months, although this is not yet
guaranteed.  We were provided a table of the expected number of supernova light
curves in each redshift bin per month, and have incorporated a total of 500
objects into our pre-JDEM priors.  We also assume that the HST servicing
mission will achieve its goals, and that HST will be used in the next 5 years
to provide a sample of supernovae beyond $z = 1$ that is 4 times as large as
the present published sample.  The Dark Energy Survey is an ambitious plan to
build a large-format camera for the 4-meter Blanco telescope at CTIO. We thank
the DES SN working group for providing us with their current projections of DES
performance in advance of a final determination of the DES strategy. Adding
their projected sample of 3000 objects, but truncating the sample beyond
redshift of 0.8 (which is a conservative assumption), we construct the
following table, which we believe represents a reasonable estimate of what will
be available for detailed cosmological analysis in 2016 (the beginning of the
JDEM mission).

Current methods for estimating distances to individual objects have an
uncertainty that is reckoned to be below 10\%, so even accounting for
photometric uncertainty, the sample sizes out to a redshift of 0.8 will be
sufficiently large that the present systematic floor will be the dominant
uncertainty in each redshift bin.  As we emphasized above, this estimate is
difficult to verify, and limits the precision that is reasonable to ascribe to a
Figure of Merit.  While most of the supernovae in this table come from DES, we
note that the sample size predicted for the Pan-STARRS program would be
sufficient to make the systematic floor into the effective limit on the distance
measurement at these redshifts.  The supernovae beyond redshift 1 will be
extremely interesting, but their statistics will not be sufficient to bring them
down to the systematic floor.  For these objects, an independent distance error
of 30\% per object has been assigned.

\section{Pre-JDEM Performance}

After constructing the Fisher matrices as described in Appendix A, we followed
the procedure outlined in Sec.\ \ref{quant}.  The results are summarized in
Table \ref{tab:werrors}.

\begin{table}
\begin{ruledtabular}
\begin{tabular}{ccccc}
                      & \rule[-2mm]{0mm}{6mm} $\,\,\, z_p\,\,\,$  
                      & \rule[-2mm]{0mm}{6mm} $\sigma(w_p)$    
                      & \rule[-2mm]{0mm}{6mm} $\sigma(w_a)$  
                      & \rule[-2mm]{0mm}{6mm} $\sigma(\Delta\gamma)$   \\ 
		      \hline\hline
\rule[-2mm]{0mm}{6mm} SNe+Planck           & 0.01 & 0.34 & 6.97 & ---  \\
\rule[-2mm]{0mm}{6mm} WL+Planck            & 0.50 & 0.09 & 1.20 & 0.51 \\
\rule[-2mm]{0mm}{6mm} \,\,SN+WL+Planck\,\, & 0.36 & 0.05 & 0.68 & 0.39 \\
\rule[-2mm]{0mm}{6mm} BAO+Planck           & 0.52 & 0.04 & 0.44 & ---  \\
\rule[-2mm]{0mm}{6mm} All                  & 0.49 & 0.03 & 0.31 & 0.21 \\
\end{tabular}
\end{ruledtabular}
\caption{Pre-JDEM expansion and growth history parameter errors.  For each
combination of probes, we show the redshift pivot values $z_p$,  68\% C.L.
Fisher matrix errors on the   DETF parameters $w_p$ and $w_a$,  and the growth
index $\gamma$.  We have projected parameters describing the  equation of state
$w(z)$ onto the $(w_p, w_a)$ subspace and, to report  these errors, fixed the
parameters $\gamma$ and $\ln G_0$ while marginalizing over the other six
parameters.  The error on  $\gamma$, on the other hand, assumes marginalization
over all other 44  parameters.
\label{tab:werrors}}
\end{table}

The FoMSWG Fisher matrices for individual techniques, as well as the techniques
in combination, can be found on the FoMSWG website {\tt
http://jdem.gsfc.nasa.gov/FoMSWG}.  Brief explanations of the format of the
Fisher matrices (which follows the format defined here) can be found there.

Also at the above website is a {\tt C} program to take the pre-JDEM Fisher
matrices as input and combine them in various ways to produce output files of
the principal components, the uncertainties in the coefficients of the principal
components, information about the DETF $\sigma(w_0)$ and $\sigma(w_a)$, and
other information.  There is also the option of including additional Fisher
matrices (say for an imagined JDEM or for a different realization of a pre-JDEM
Fisher matrix). A guide to the software tools is described in the next section.

\subsection{Practical Guide to the FoMSWG Software}

In association with this publication, we provide some basic software
to help perform the analyses outlined herein. The software package is
available from the JDEM website as a gzipped tarball (named {\tt
public\_fomswg\_code.tar.gz}) and contains the following items in the
main directory {\tt /PUBLIC\_CODE}:

\begin{itemize}

\item {\tt fisher\_fomswg.c}, which is the main C program for computing the
  Principal Components and their accuracies. This uses standard C libraries.

\item {\tt nrutil.c}, which provides some Numerical Recipes
  \cite{NumericalRecipes} utilities adapted for double precision.

\item {\tt function.c}, which contains specific functions for the manipulation
  of matrices.

\item The {\tt /DATA} sub-directory which contains the matrices discussed in
  Sec.\ \ref{Format} for Planck and the pre--JDEM BAO, SN and WL observations
  (see Appendix A for a detailed explanation of the assumptions that when into
  constructing these matrices). Note that some matrices have maximal indices
  less than 44 or are missing a few lower indices; this is simply due to the
  fact that the technique is insensitive to the parameter in question (as listed
  in Sec.\ \ref{Format}). The associated Fisher matrix elements will thus be
  zero (explicitly listing these elements as zeros in data-files is equivalent
  and also works).

\item A compiled executable, {\tt fish}, is provided and may work immediately
  on a linux-based system.  If not, a {\tt makefile} is provided which should
  allow the code to be compiled on most UNIX/Linux systems (we have compiled
  the code under Mac OSX, Debian, SuSE and IRIX).  To compile the code, remove
  the old {\tt fish} executable and create a new one by typing {\tt make fish}.

\end{itemize}

The {\tt fish} program can be run from the command line using the default
matrices given in the {\tt /DATA} sub-directory, or the user can input a
customized matrix (as long as it follows our conventions). To study multiple
techniques of the user's choosing, the user should add the respective Fisher
matrices and input only the single, combined matrix. 

In default mode, the user can select from eight possible combinations of SN,
BAO, WL, and Planck, and the {\tt fish} program provides a variety of numerical
outputs, which are stored in the {\tt /OUTPUTS} sub-directory and displayed on
screen. 

The screen outputs include:

\begin{itemize}

\item Errors on the parameters $w_0$ and $w_a$, marginalized over all other 9
parameters.

\item DETF FoM (defined as $\left[\sigma(w_a)\times\sigma(w_p)\right]^{-1}$).

\item Pivot redshift ($z_p$) and the uncertainty in the associated pivot
equation of state ($w_p$),  see Eq.\ (\ref{wp}).

\item Error in constant $w$, marginalized over all other 9 parameters.

\item The growth index, $\gamma$, and its FoM [defined as
$\sigma^{-2}(\Delta\gamma)$] if the user selects options including WL
measurements (options 2, 4, 5, 8). 

\end{itemize}

If the user picks options (1, 3, 6, 7) that only include combinations of SN,
BAO and Planck (i.e., geometrical measurements), then $\gamma$ and $G_0$ are
fixed and the following file outputs are provided:

\begin{itemize}

\item {\tt sigma.dat}, which contains the errors on the 36 Principal
Components (with and without priors).

\item {\tt PC\_1234.dat}, which contains the first four eigenvectors.

\item {\tt PC\_all.dat}, which contains all 36 eigenvectors. Both are provided
as a function of redshift (36 bins) as discussed in Sec. \ref{Modeling}.

\item {\tt w0wa.dat}, which is a $2\times2$ Fisher matrix for the $w_0$
and $w_a$ parameterization.

\end{itemize}

For other options, which include the WL measurements (options 2, 4, 5, 8),  the
program provides an extra set of files (in addition to the ones listed above)
which allow $\gamma$ (and $G_0$) to vary. These files have the same naming
scheme as above, but with an additional {\tt \_gamma} phrase attached, e.g.,
{\tt sigma\_gamma.dat}.

The software is provided to help people become accustomed with the analyses
outlined herein, but has not been rigorously tested, and the user is urged not
to use its outputs blindly without checking them. Comments on the code should
be sent to Dragan Huterer, who deserves the credit for producing this tool.




\begin{thebibliography}{99}
\frenchspacing

\bibitem{REVIEWS}
Two recent reviews of dark energy are
E. J. Copeland, M. Sami and S. Tsujikawa,
Int. J. Mod. Phys. D {\bf 15}, 1753 (2006),
and
J. A. Frieman, M. S. Turner, and D. Huterer,
Ann. Rev Astron. and Astrop. {\bf 46}, 385 (2008).

\bibitem{Ptolemy}  
See, e.g., G. J. Toomer, 
{\it Ptolemy's Almagest} (Princeton University Press, 1998).

\bibitem{Bousso:2000xa}
R. Bousso and J. Polchinski,
JHEP {\bf 0006}, 006 (2000),

\bibitem{Kachru:2003aw}
S. Kachru, R. Kallosh, A. Linde and S. P. Trivedi,
Phys. Rev. D {\bf 68}, 046005 (2003).

\bibitem{Bousso:2007gp}
R. Bousso,
Gen. Rel. Grav.  {\bf 40}, 607 (2008).

\bibitem{Dyson:2002pf}
L. Dyson, M. Kleban and L. Susskind,
JHEP {\bf 0210}, 011 (2002)

\bibitem{Albrecht:2004ke}
A. Albrecht and L. Sorbo,
Phys. Rev.  D {\bf 70}, 063528 (2004)

\bibitem{Efstathiou:1990xe}
G. Efstathiou, W. J. Sutherland and S. J. Maddox,
Nature {\bf 348}, 705 (1990).

\bibitem{Barnard:2008mn}
M. Barnard, A. Abrahamse, A. Albrecht, B. Bozek and M. Yashar,
Phys. Rev.  D {\bf 78}, 043528 (2008)

\bibitem{fr}
For a review, see 
L. Amendola, R. Gannouji, D. Polarski and S. Tsujikawa,
Phys. Rev. D {\bf 75}, 083504 (2007).

\bibitem{braneworld}
For a braneworld review, see
R. Maartens,
Living Rev. Rel. {\bf 7}, 7 (2004).
There are many braneworld modifications that are possible explanations of dark
energy, or at least modify gravity.  For example, see,
P. Binetruy, C. Deffayet, U. Ellwanger and D. Langlois,
Phys. Lett. {\bf B 477}, 285 (2000);
C. Deffayet, G. R. Dvali and G. Gabadadze,
Phys. Rev. D {\bf 65}, 044023 (2002);
R. Gregory, V. A. Rubakov and S. M. Sibiryakov,
Phys. Lett. {\bf B 489}, 203 (2000);
G. R. Dvali, G. Gabadadze and M. Porrati,
Phys. Lett. {\bf B 485}, 208 (2000);
C. Csaki, J. Erlich, T. J. Hollowood and J. Terning,
Phys. Rev. D {\bf 63}, 065019 (2001);
I. I. Kogan, S. Mouslopoulos, A. Papazoglou, G. G. Ross and J. Santiago,
Nucl. Phys. {\bf B 584}, 313 (2000).

\bibitem{DETF} 
A. Albrecht {\it et al.}, 
{\it Report of the Dark Energy Task Force,}
arXiv:astro-ph/0609591.

\bibitem{Eisenstein2007}
D. J. Eisenstein, H. Seo, E. Sirko, and D. N. Spergel, 
Astrophys. J. {\bf 664} 675 (2007).

\bibitem{Rosati} 
P. Rosati, S. Borgani, and A. Norman,
Ann. Rev Astron. and Astrop. {\bf 40}, 539 (2002).

\bibitem{Rykoff}
E. S. Rykoff {\it et al.}, 
Mon. Not. Roy. Astron. Soc. {\bf 387}, L28 (2008).

\bibitem{Pratt}
G. W. Pratt, J. H. Croston, M. Arnaud and H. Boehringer,
{\it Galaxy cluster X-ray luminosity scaling relations from a representative 
local sample (REXCESS),}
arXiv:0809.3784 [astro-ph].

\bibitem{Kravtsov}
A. V. Kravtsov, A. Vikhlinin, and D. Nagai, 
Astrophys. J. {\bf 650}, 1 (2006).

\bibitem{Arnaud}
M. Arnaud, E. Pointecouteau and G. W. Pratt, 
Astronomy and Astrophysics {\bf 474}, L37 (2007).

\bibitem{Cohn}
J. D. Cohn, A. E. Evrard, M. White, D. Croton and E. Ellingson,
Mon. Not. Roy. Astron. Soc. {\bf 382}, 1738 (2007).

\bibitem{MM2004} 
S. Majumdar and J. J. Mohr,
Astrophys. J. {\bf 613}, 41 (2004).

\bibitem{Lima_Hu}
M. Lima and W. Hu, 
Phys. Rev. D {\bf 70}, 043504 (2004); Phys. Rev. D {\bf 72}, 043006 (2005).

\bibitem{Schuecker}
P. Schuecker {\it et al.}, 
Astron and Astrop. {\bf 398}, 867 (2003).

\bibitem{Allen}
S. W. Allen {\it et al.}, 
Mon. Not. Roy. Astron. Soc. {\bf 383}, 879 (2008).

\bibitem{Vikhlinin}
A. Vikhlinin {\it et al.}, 
{\it Chandra Cluster Cosmology Project II: Samples and X-ray Data Reduction,}
arXiv:0805.2207 [astro-ph].

\bibitem{Johnston:2007uc}
D. E. Johnston {\it et al.}  [SDSS Collaboration],
{\it Cross-correlation Weak Lensing of SDSS galaxy Clusters II: Cluster Density
Profiles and the Mass--Richness Relation,}
arXiv:0709.1159 [astro-ph].

\bibitem{Guzzo08} 
L. Guzzo {\it et al.}, 
Nature {\bf 451}, 541 (2008).

\bibitem{Kaiser87} 
N. Kaiser,  
Mon. Not. Roy. Astron. Soc.  {\bf 227}, 1 (1987).

\bibitem{Wang08} 
Y. Wang,
JCAP {\bf 05}, 021 (2008).

\bibitem{Linder08} 
E. Linder, 
Astropart. Phys. {\bf 29}, 336 (2008).

\bibitem{Hamilton98} 
A. J. Hamilton, 
in {\it The Evolving Universe,} ed. D. Hamilton, (Kluwer, p. 185, 1998). 

\bibitem{Aquaviva08}
V. Aquaviva, A. Hajian, D. Dn. Spergel, S. Das,
Phys. Rev. D {\bf 78}, 043514 (2008). 

\bibitem{SongPercival08} 
Y.-S. Song, and W. J. Percival,
{\it Reconstructing the History of Structure Formation using Peculiar 
Velocities,}
arXiv:0807.0810 [astro-ph].
 
\bibitem{PercivalWhite08} 
W. Percival and M. White, 
{\it Testing cosmological structure formation using redshift-space distortions,}
arXiv:0808.0003 [astro-ph].

\bibitem{2002MNRAS.335..432V}
L. Verde, {\it et al.,}
Mon. Not. Roy. Astron. Soc. {\bf 335}, 432 (2002).

\bibitem{McDonald_Seljak08} 
P. Mc Donald and U. Seljak, 
{\it How to measure redshift-space distortions without sample variance,}
arXiv:0810.0323 [astro-ph].

\bibitem{White08} 
M. White, Y.-S. Song, and W. J. Percival,
{\it Forecasting Cosmological Constraints from Redshift Surveys,}
arXiv:0810.1518 [astro-ph].
  
\bibitem{1967ApJ...147...73S}
R. K. Sachs and A. M. Wolfe, 
Astrophys. J. {\bf 147}, 73 (1967).

\bibitem{1996PhRvL..76..575C}
R. G. Crittenden and N. Turok,
Phys. Rev. Lett. {\bf 76}, 575 (1996).

\bibitem{2008PhRvD..78d3519H} 
S. Ho, C. Hirata, N. Padmanabhan, U. Seljak and N. Bahcall,
Phys. Rev. D {\bf 78}, 043519 (2008).

\bibitem{2008PhRvD..77l3520G}
T. Giannantonio, R. Scranton, R. G. Crittenden, R. C. Nichol, S. P. Boughn, 
A. D. Myers and G. T. Richards,
Phys. Rev. D {\bf 77}, 123520 (2008).

\bibitem{2004PhRvD..70h3536A}
N. Afshordi,
Phys. Rev. D {\bf 70}, 083536 (2004).

\bibitem{AP79}
C. Alcock and B. Paczynski, 
Nature {\bf 281}, 358 (1979).

\bibitem{Ballinger96}
W. E. Ballinger, J. A. Peacock, and A. F. Heavens, 
Mon. Not. Roy. Astron. Soc. {\bf 282}, 877 (1996).

\bibitem{Matsubara02}
T. Matsubara and A. S. Szalay, 
Astrophys. J. {\bf 556}, L67 (2001).

\bibitem{Shoji08}
M. Shoji, D. Joeng, and E. Komatsu,
{\it Extracting Angular Diameter Distance and Expansion Rate of the Universe
from Two-dimensional Galaxy Power Spectrum at High Redshifts: Baryon Acoustic
Oscillation Fitting versus Full Modeling,}
arXiv:0805.4238 [astro-ph].

\bibitem{Ross}
N. Ross, et al., 
Mon. Not. Roy. Astron. Soc. {\bf 381}, 573 (2007).

\bibitem{Hut_Stark}
D. Huterer and G. Starkman,
Phys. Rev. Lett. {\bf 90}, 031301 (2003).

\bibitem{Linder_2005}
E. V. Linder, Phys. Rev. D {\bf 72}, 043529 (2005).
  
\bibitem{2006astro.ph..4069T}
The Planck Collaboration,
{\it Planck: The scientific programme,}
arXiv:astro-ph/0604069.

\bibitem{2007ApJS..170..335P}
L. Page, {\it et al.}, 
Astrophys. J. Suppl. {\bf 170}, 335 (2007).

\bibitem{2003ApJ...595...13H}
G. Holder, Z. Haiman, M. Kaplinghat and L. Knox,
Astrophys. J. {\bf 595}, 13 (2003).

\bibitem{EisensteinHu}
D. J. Eisenstein, and W. Hu, 
Astrophys. J. {\bf 496} 605 (1998).

\bibitem{Smithetal}
R. E. Smith {\it et al.}  [The Virgo Consortium Collaboration],
Mon. Not. Roy. Astron. Soc. {\bf 341}, 1311 (2003).

\bibitem{2001ApJ...554...56C}
A. Cooray and W. Hu,
Astrophys. J. {\bf 554}, 56 (2001).

\bibitem{1999MNRAS.308..119S}
R. K. Sheth and G. Tormen,
Mon. Not. Roy. Astron. Soc. {\bf 308}, 119 (1999).

\bibitem{1997ApJ...490..493N}
J. F. Navarro, C. S. Frenk and S. D. M. White,
Astrophys. J. {\bf 490}, 493 (1997).

\bibitem{2003ApJ...593..272V}
G. M. Voigt {\it et al.}, 
Astrophys. J. {\bf 593}, 272 (2003).

\bibitem{2007MNRAS.376...13M} 
R. Massey {\it et al.},
Mon. Not. Roy. Astron. Soc. {\bf 376}, 13 (2007).
 
\bibitem{2003MNRAS.343..459H}
C. M. Hirata and U. Seljak,
Mon. Not. Roy. Astron. Soc. {\bf 343}, 459 (2003).

\bibitem{2008ApJ...684...88N}
J. A. Newman,
Astrophys. J. {\bf 684}, 88 (2008).

\bibitem{2003PhRvL..91n1302J}
B. Jain and A. Taylor,
Phys. Rev. Lett. {\bf 91}, 141302 (2003).

\bibitem{2004ApJ...600...17B}
G. M. Bernstein and B. Jain,
Astrophys. J. {\bf 600}, 17 (2004).

\bibitem{2007NJPh....9..444B}
S. Bridle and L. King,
New J. Phys. {\bf 9}, 444 (2007).

\bibitem{2004ApJ...601L...1T}
M. Takada and M. J. White,
Astrophys. J. {\bf 601}, L1 (2004).

\bibitem{2007MNRAS.381.1197H}
C. M. Hirata {\it et al.},
Mon. Not. Roy. Astron. Soc.{ \bf 381}, 1197 (2007).

\bibitem{2006MNRAS.371..750H}
C. Heymans {\it et al.},
Mon. Not. Roy. Astron. Soc. {\bf 371},  750 (2006).
	
\bibitem{Seoa}
H. Seo and D. J. Eisenstein,
Astrophys. J. {\bf 633}, 575 (2005).

\bibitem{Seob}
H. Seo, E. R. Siegel, D. J. Eisenstein, and M. White, 
Astrophys. J. {\bf 686}, 13 (2008).

\bibitem{Sanchez}
A. G. S\'anchez, C. M. Baugh, and R. Angulo, 
Mon. Not. Roy. Astron. Soc. {\bf 390}, 1470 (2008).

\bibitem{Conley.2007}
A. Conley {\it et al.},
Astrophys. J. {\bf 664}, L13 (2007).

\bibitem{Wood-Vasey.2007b}
M. Wood-Vasey {\it et al.},
{\it Type Ia Supernovae are Good Standard Candles in the Near Infrared: Evidence
from PAIRITEL,}
arXiv:0711.2068 [astro-ph].

\bibitem{Krisciunas.2007}
K. Krisciunas {\it et al.},
Astron. J. {\bf 133}, 58. (2007)

\bibitem{Riess.1998}
A. Riess {\it et al.},
Astron. J. {bf\ 116}, 1009 (1998). 

\bibitem{Perlmutter.1999}
S. Perlmutter {\it et al.},
Astrophys. J. {\bf 517}, 565 (1999).

\bibitem{Wood-Vasey.2007a}
M. Wood-Vasey {\it et al.},
Astrophys. J. {\bf 666}, 716 (2007).

\bibitem{Astier.2006}
Astier, P. {\it et al.},
Astronomy and Astrophysics {\bf 447}, 31 (2006).

\bibitem{Riess.2004}
A. Riess {\it et al.},
Astrophys. J. {\bf 607}, 665 (2004).

\bibitem{Riess.2007}
A. Riess {\it et al.},
Astrophys. J. {\bf 659}, 98 (2007).

\bibitem{Jha.2007}
S. Jha, A. Riess, and R. Kirshner
Astrophys. J. {\bf 659}, 122 (2007).

\bibitem{Kowalski.2008}
M. Kowalski {\it et al.},
Astrophys. J. {\bf 686}, 749 (2008).

\bibitem{NumericalRecipes} 
W. H. Press, S. A. Teukolsky, W. T. Vetterling, and B. P. Flannery,
{\it Numerical Recipies in C: The Art of Scientific Computing: Second Edition 
(1992)}, (Cambridge University Press, Cambridge, 1992).
 

\end{thebibliography}
\end{document}